\documentclass{aa}
\usepackage[varg]{txfonts} 

\usepackage{amsmath}
\usepackage{siunitx}
\usepackage{xfrac}
\usepackage{natbib}
\usepackage[table,x11names]{xcolor}
\usepackage{graphicx}
\usepackage{booktabs}
\usepackage{csquotes}
\usepackage{appendix} 
\usepackage{subcaption}
\usepackage{multirow,array}
\usepackage{bbding}
\usepackage{amsmath}
\usepackage[T1]{fontenc}
\usepackage{pifont}
\makeatletter
  \renewcommand*\aa@pageof{, page \thepage{} of \pageref*{LastPage}}
\makeatother
\usepackage{hyperref}

\hypersetup{pdfpagemode = {UseNone},
            pdftitle = {Resonant overstability in hydrosimulations},
            pdfcreator = {\LaTeX},
            pdfproducer = {pdfeTeX-0.\the\pdftexversion\pdftexrevision},
            pdfauthor = {Zahra Afkanpour, Sareh Ataiee},
            pdfsubject = {},
            pdfview = {FitH},
            pdfstartview = {FitH},
            colorlinks = {true},
            linkcolor = [rgb]{0,0.35,0.7},
            citecolor = [rgb]{0,0.35,0.7},
            filecolor = [rgb]{0.61,0,0},
            urlcolor = [rgb]{0.61,0,0},
           }

\usepackage[normalem]{ulem}
\usepackage{enumitem}

\usepackage{bm}

\colorlet{darkpink}{pink!50!black}
\colorlet{darkgreen}{green!50!black}

\definecolor{maroon}{rgb}{0.69, 0.19, 0.38}
\newcommand{\refn}[1]{\textcolor{maroon}{\bf (REF NEEDED)}}

\newcommand{\beq}{\begin{equation}}
\newcommand{\eeq}{\end{equation}}

\begin{document}

\title{Overstability of the 2:1 mean motion resonance: Exploring disc parameters with hydrodynamic simulations}

\author{Zahra Afkanpour\inst{\ref{inst1},\ref{inst2}},
		Sareh Ataiee\inst{\ref{inst1}},
		Alexandros Ziampras \inst{\ref{inst3}},
		Anna B. T. Penzlin\inst{\ref{inst2},\ref{inst4}},
		Rafael Sfair\inst{\ref{inst2},\ref{inst5}},
		Christoph Sch\"afer\inst{\ref{inst2}},
		Wilhelm Kley\inst{\ref{inst2}}\thanks{W.~Kley has passed away but is included as a co-author for his guidance during this project.},
		Hilke Schlichting\inst{\ref{inst6}}
		}

\institute{
Department of Physics, Faculty of Sciences, Ferdowsi University of Mashhad, Mashhad, 91775-1436, Iran \label{inst1}
\and
Institut f\"ur Astronomie und Astrophysik, Universität T\"ubingen, Auf der Morgenstelle 10, 72076 T\"ubingen, Germany\label{inst2}
\and
Astronomy Unit, School of Physical and Chemical Sciences, Queen Mary University of London, London E1 4NS, UK\label{inst3}
\and
Astrophysics Group, Department of Physics, Imperial College London, Prince Consort Rd, London, SW7 2AZ, UK \label{inst4}
\and
Grupo de Dinâmica Orbital e Planetologia, São Paulo State University, UNESP, Guaratinguetá, CEP 12516-410, São Paulo, Brazil \label{inst5}
\and
Department of Earth, Planetary, and Space Sciences, The University of California, Los Angeles, 595 Charles E. Young Drive East, Los
Angeles, CA 90095, USA \label{inst6}
\\
\email{zahra.afkanpour@mail.um.ac.ir,
	 sarehataiee@um.ac.ir}\\
}

\date{}

\abstract
{
\textit{Context.} Resonant planetary migration in protoplanetary discs can lead to an interplay between the resonant interaction of planets and their disc torques called overstability. While theoretical predictions and N-body simulations hinted at its existence, there was no conclusive evidence until hydrodynamical simulations were performed.\\
\textit{Aims.} Our primary purpose is to find a hydrodynamic setup that induces overstability in a planetary system with two moderate-mass planets in a first-order 2:1 mean motion resonance. We also aim to analyse the impact of key disc parameters, namely the viscosity, surface density, and aspect ratio, on the occurrence of overstability in this planetary system when the masses of the planets are kept constant.\\
\textit{Methods.} We performed 2D locally isothermal hydrodynamical simulations of two planets, with masses of 5 and 10 $M_{\oplus}$, in a 2:1 resonance. Upon identifying the fiducial model in which the system exhibits overstability, we performed simulations with different disc parameters to explore the effects of the disc on the overstability of the system.\\
\textit{Results.} We observe an overstable planetary system in our hydrodynamic simulations. In the parameter study, we note that overstability occurs in discs characterised by low surface density and low viscosity. Increasing the surface density reduces the probability of overstability within the system. A limit cycle was observed in a specific viscous model with $\alpha_{\nu} = 10^{-3}$. In almost all our models, planets create partial gaps in the disc, which affects both the migration timescale and structure of the planetary system.\\
\textit{Conclusions.} We demonstrate the existence of overstability using hydrodynamic simulations but find deviations from the analytic approximation and show that the main contribution to this deviation can be attributed to dynamic gap opening.\\

}

\titlerunning{Overstability at 2:1 resonance}
\authorrunning{Afkanpour et al.}

\maketitle

\section{Introduction}\label{sec:intro}

Two key questions raised by the rich body of exoplanet data provided by the \textit{Kepler} mission were why most exoplanets in multi-planet systems are not in mean motion resonance~(MMR) and why there are pileups just wide of exact commensurabilities ~\citep{2014ApJ...790..146F}. Identifying physical processes that could explain the rarity of exoplanets at exact MMRs when there is a sudden increase in the planets' number at slightly larger period ratios sparked numerous investigations \citep[e.g.][]{2013ApJ...778....7B, 2013ApJ...770...24P, 2014A&A...570L...7D,2014AJGoldreich,2017AJ....154..236W,2022MNRAS.514.3844C}. One pioneering study was carried out by \citet{2014AJGoldreich}, who suggested the majority of observed exoplanets cannot permanently stay in an MMR because their masses are in a regime such that even if they are captured in an MMR, their orbits become overstable in a time comparable to their eccentricity damping timescale ($\tau_{e}$) and the resonance is eventually broken\footnote{We frequently use the word 'resonance' instead of MMR unless stated otherwise.}. They conclude that since $\tau_{e}$ is typically much shorter than the migration timescale, $\tau_{a}$ \citep[e.g.][]{1980ApJ...241..425G}, the planets spend more time between the resonances rather than in them. This finding triggered many follow-up studies.

\citet{2014AJGoldreich} analytically investigated the stability of a two-planet system undergoing convergent migration in the presence of planet eccentricity damping by the disc. Assuming the outer planet is more massive than the inner one (a restricted three-body problem) and that their eccentricities are very small, they found three possible fates for the system. Depending on the mass of the outer planet, the system can (a)~be captured permanently in resonance with a constant resonance angle and period ratio (stability), (b)~stay in resonance while the resonance angle and the period ratio oscillate with a constant amplitude (limit cycle), or (c)~be caught in resonance for a time of about $\tau_{e}$ but eventually become overstable. Overstability happens when the amplitudes of the oscillating eccentricities and resonance angles grow until the planetary orbits circulate and the planets come out of the resonance. \citet{2014AJGoldreich} analytically calculated the criterion for overstable librations and show that it depends on the eccentricity damping timescale, migration timescale, and planet-to-star mass ratio when the restricted three-body problem applies. \citet{2015A&A...579A.128D} and \citet{2015Deck} improved upon the study of \cite{2014AJGoldreich} by taking the masses of both planets into account. They show that the ratio of the two planets' masses is a key parameter when determining the outcome of a system with convergent migration and the time spent in resonance when overstable librations lead to escape. \citet{2015Deck} categorised the outcome of the system into three regimes, similar to \citet{2014AJGoldreich} but allowed for a finite, and in fact often small, mass ratio between the two planets. They also studied the long-term evolution of the systems and show that for a planetary pair undergoing overstability, the escape time from the resonance can be 5--10~times longer than $\tau_{e}$. Moreover, an overstable system goes through successive resonance capturing, with several consequences. Firstly, because the subsequent resonances are more compact, if the system is captured in a stable resonance, the final spacing of the system would be very closely packed. However, if the next resonances are overstable, the system cascades until it reaches the chaotic zone where various MMRs overlap, and the planets are scattered. Secondly, because an overstable system stays in each subsequent resonance for a time longer than $\tau_{e}$, the system would spend most of its evolutionary time in resonances rather than between them. These results hint that using more realistic models probably leads to even longer staying times in resonances for the system.

One step towards more realistic models is to perform N-body simulations so that the gravitational interaction between the objects can be calculated with fewer simplifications. \citet{2015MNRAS.449.3043X} performed a numerical survey to study the long-term evolution of co-planar low-mass two-planet systems near the disc inner edge, which guarantees the convergent migration of the planets. They included the effect of disc--planet tidal interaction with two acceleration terms in the equation of motion of the planets. These terms depend on $\tau_{e}$ and $\tau_{a}$, which were considered constant during the whole evolution time unless the planets reached the disc inner edge. They find that only a few percent of their models ended in second-order resonances, some of which were overstable. Later, \citet{2018MNRAS.481.1538X} studied first-order MMR capturing using N-body simulations and taking the eccentricity-related non-linear terms into account when determining $\tau_{e}$ and $\tau_{a}$ \citep{2008A&A...482..677C}. They conclude that, compared to models with constant timescales, the eccentricity of the captured planets is larger and the resonant systems tend to be more stable. Recently, \citet{2022ApJ..Nesvorny} conducted a study on the exoplanetary system TOI-216, finding that a large eccentricity and a 2:1 resonance libration amplitude of the inner planet suggest the presence of a limit cycle during the disc phase. These results highlight the need to investigate the overstability of resonant systems using hydrodynamical simulations.

Hydrodynamical modelling of low-mass resonant planetary systems is very computationally expensive because: (a)~a very long running time is needed to bring the planets into resonance and to follow the stability of the system afterwards, and (b)~such simulations need high spatial resolution to properly resolve the disc torques on the planets. Among the attempts at modelling the low-mass resonant systems using hydrodynamical simulations \citep[e.g.][]{2005MNRASPapaloizou,2008A&APierens,2011A&A...531A...5P,2013MNRAS.434.3018P,2018MNRASHands,2021A&AAtaiee}, overstability has only been reported in \citet{2018MNRASHands} and \citet{2021A&AAtaiee}. In the former, the orbital evolution of two moderate-mass planets in a 2:1 MMR in a locally isothermal disc was studied for various disc viscosities using the \texttt{PLUTO} code\footnote{http://plutocode.ph.unito.it}, as well as the \texttt{FARGO3D} code\footnote{http://fargo.in2p3.fr} for one test model. In all but one of their models, the planets cross the resonance; the exception is a model run with \texttt{FARGO3D} and described in their appendix. In all the other  models, while the \citet{2014AJGoldreich} overstability condition is marginally satisfied, the planets' eccentricities are excited as the planets cross the 2:1 MMR and are damped afterwards. However, in the model in their appendix, the planets stay in the resonance for more than \num{2000}~years; the oscillation amplitude of the eccentricities slowly grows until the resonance is broken. This is the typical behaviour of an overstable system \citep[e.g.][]{2014AJGoldreich,2015Deck}. Recently, \citet{2021A&AAtaiee} found some overstable systems in their study of moderate-mass resonant multi-planet systems close to a disc inner edge. They performed locally isothermal hydrodynamical simulations using the \texttt{FARGO} code\footnote{http://fargo.in2p3.fr/-Legacy-archive-} and modelled the inner disc edge as either smooth or sharp as the surface density drops. Only for a smooth inner edge with a small disc aspect ratio of \num{0.03} did they find overstable systems that clearly cascade into higher-order packed resonances. These cases suggest that overstability can indeed happen in more realistic models. The notable point about these two hydrodynamical studies is that the planets are not low-mass planets that undergo type-I migration without opening gaps but instead are moderate-mass planets that open partial gaps.

In hydrodynamical simulations, the planets' eccentricity damping, migration, and disc structure are all interrelated, in contrast to analytical studies and N-body simulations. When the planets open a (partial)~gap, their eccentricity damping and migration timescales are modified. Moreover, the planetary gaps can merge as the planets' orbits approach each other. This can create asymmetries in the torques on the planets \citep{2001MNRAS.320L..55M} and cause eccentricity damping. The question of how the eccentricity of a gap-opening planet is damped is answered in \citet{2023A&A...Pichierri}. Another important issue that is automatically included in hydrodynamical simulations is the time evolution of $\tau_{a}$ and $\tau_{e}$. They are two crucial parameters in forming overstable systems and are undoubtedly time-dependent, especially if the disc viscosity is low. As a planet migrates and/or opens a (partial)~gap, the torque on the planet that determines these two timescales changes \citep{2014MNRAS.444.2031P,2019MNRAS.484..728M}. Whether the criteria suggested by the analytical calculations are valid and how gap opening affects overstability are questions that need further investigation.

In this study we aim to inspect the overstability at 2:1 MMR for a system with two moderate-mass planets. Because of the complexity of the models in \citet{2021A&AAtaiee} due to the inner-edge surface density evolution, we started with the model in the appendix of \citet{2018MNRASHands}, determined the proper initial condition, and performed a parameter study on the disc parameters. We find various outcomes of overstability, limit cycles, and even divergent migration for different disc parameters when the planetary masses remain untouched.

The manuscript is ordered as follows. In Sect.~\ref{sec:model} we introduce our nomenclature to avoid any ambiguity. Then we detail our method, initial conditions, and setup. The results are described in Sect.~\ref{sec : results}. Thereafter, we discuss our findings and compare them with previous work in Sect.~\ref{sec: discussion}. Finally, the summary and conclusions come in Sect.~\ref{sec: conclusion}.

\section{Models}\label{sec:model}
In this section we define some frequently used key terms in the manuscript. Then, we determine the appropriate physical model and initial conditions for our fiducial model and our parameter study. Finally, we introduce the hydrodynamic codes utilised for running the simulations, along with details about the setup.

\subsection{Nomenclature} \label{subsec:defs}
We define several commonly used terms in the manuscript as follows:

\begin{description}[leftmargin=0pt,font=\itshape]
\item[Commensurability:] A commensurability is where the period (or mean motion) ratio of two planets equals the ratio of two integers.
\item[Mean motion resonance (MMR):] Two planets are in MMR when, in addition to being in a commensurability, at least one resonance angle stays limited to a span smaller than $2\pi$.
\item[Stability:] We call a system stable in an MMR if the planets' period ratio, eccentricities, and resonance angle(s) remain bound around an equilibrium value until the end of the simulation. It means that even if these quantities oscillate but their oscillation amplitudes remain constant over time, we consider the system stable.
\item[Limit cycle:] A stable system is in limit cycle if the planets' period ratio, eccentricities, and resonance angle(s) oscillate around an equilibrium value with constant amplitude.
\item[Overstability:] A system becomes overstable if during the captured time in MMR, the planets' period ratio, eccentricities, and resonance angle(s) oscillate around an equilibrium value with growing amplitudes until the planets' orbits circulate and the system comes out of the MMR. Such a system can be subsequently captured in either the next stable or unstable MMRs.
\item[Resonance passage:] When the planets pass an MMR location without being captured, the planets' eccentricities increase suddenly and the pattern of resonance angle changes slightly but the system does not remain in this configuration and crosses it quickly.  We call this state `resonance passage'.
\end{description}
\subsection{Physical model and initial conditions} \label{subsec: model}
Power laws are employed to describe the radial dependence of surface density $(\Sigma)$, temperature $(T)$, and scale height $(H)$ of our gaseous disc. These profiles are as follows, 
\begin{equation} \label{eq:surface density, scale height}
\begin{aligned}
\Sigma (r) &= \Sigma_0 \left(\dfrac{r}{r_0}\right)^{-\alpha_{\Sigma}}, \\
T (r) &= T_0 \left(\frac{r}{r_0}\right)^{-\beta}, \\
H(r) &= r\,h(r) = r\,h_0 \left(\dfrac{r}{r_0}\right)^{f},
\end{aligned} 
\end{equation} 

where $r_0$ represents the unit of length chosen to be 1 \textit{au} in this study. The variables $\Sigma_{0}$, $T_{0}$, and $h_{0}$ denote the surface density, temperature, and aspect ratio at $r_0$, respectively. The time of one orbit at $r_0$ was defined as one year. We assumed that the protoplanetary disc is locally isothermal. Since sound speed $c_{\rm s}$ is proportional to the square of temperature and aspect ratio is defined as $h(r)=c_{\rm s}/v_{\rm K}$ with $ v_{\rm K} $ being the Keplerian velocity, the relation between disc flaring index and temperature slope becomes $ \beta = -2f+1 $. In this study, the prescription for viscosity in the disc is the $ \alpha $-viscosity  model, which is expressed as $\nu = \alpha_{\nu} c_{\rm s} H$ \citep{1973alphavisc}. To ensure the viscous equilibrium of the system, we selected the initial condition such that the mass accretion rate through the disc remains constant, denoted as $ \dot{M} = 3 \pi \nu \Sigma $. At a unit length of $r_0 = 1$ \textit{au}, its value is approximately $\dot{M} (r = r_0) \approx 1.7 \times 10 ^{-10} $\textit{$ M_{\star}/yr $}. This condition requires that $ \alpha_{\Sigma} = 2 f + 1/2 $ in the context of an $ \alpha $-viscosity model. In all models in this study, we adopted an $\alpha_{\Sigma}$ value of 1.0, which dictates a flaring index of $f = 0.25$ for a disc in viscous equilibrium.

In the fiducial model, we set $\Sigma_0$ to $1.13 \times 10^{-4}\,M_{\star}/r_{0}^{2}$, which corresponds to \num{1000}\,g/cm${}^2$ for a host star with a solar mass,  $M_{\star} = 1\,\mathrm{M}_\odot$. In addition, we set $h_0$ to \num{0.05} and used $10^{-5}$ for $\alpha_{\nu}$. 

In our parameter study, we investigated the occurrence of overstability by varying the parameters of the disc. Specifically, we made changes to $ \Sigma_{0} $, $ \alpha_{\nu} $, and $ h_{0} $. We adjusted the value of $ \Sigma_{0} $ to $[1/4, 1/3, 1/2, 2, 3]$ times the value in the fiducial model. We also examined the values of $ \alpha_{\nu} \in [10^{-6}, 10^{-4}, 10^{-3}]$. For $ h_{0} $, we considered two smaller and one larger values as \num{0.03}, \num{0.04}, and \num{0.06}.

Inspired by the study conducted by \citet {2018MNRASHands}, our models consist of two super-Earth mass planets initially in circular orbits with no inclination. Accretion onto the planets is not considered. The mass ratio of the inner planet to the star is $q_1 = 1.5\times 10^{-5}$, and the mass ratio of the outer planet to the star is $q_2 = 3\times 10^{-5}$. This corresponds to \num{5} and \num{10} times the Earth's mass ($M_{\oplus}$), respectively, assuming the host star has a solar mass. These planets are positioned near 2:1 resonance at orbital radii of $ a_{1} = 1.23\ r_0 $ and $ a_{2} = 2.0\ r_0 $ to save the computational cost. The planets immediately begin migrating, and depending on the specific model, they may eventually reach the 2:1 resonance.

\subsection{Codes and setup} \label{subsec: codes}
In our numerical study, we utilised two different codes to conduct our research. The main code we employed was \texttt{FARGO3D}, developed by \citet{2016ApJS.Benitez}. This code allows us to solve the hydrodynamics equations with a vertically integrated approach. To directly compare our results with a previous study by \citet{2018MNRASHands}, we also performed simulations of the fiducial model using the \texttt{PLUTO} code \citep{2007Pluto}.

\begin{figure}
	\centering
	\resizebox{\hsize}{!}{\includegraphics{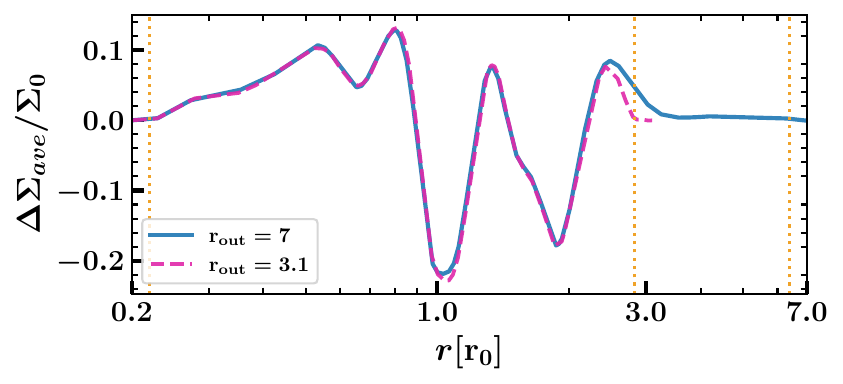}}
	\caption{Azimuthally averaged surface density perturbation at $t = 10^4$ years for the fiducial model with $r_\mathrm{out} = 7 r_0$ (solid blue line) and \num{3.1}$r_0$, as in HA18 (dotted magenta). The vertical lines show the damping radius. The damping zones were placed between [0.2--0.22]$r_0$ and [2.82--3.1]$r_0$ for a model with $r_\mathrm{out} = 3.1 r_0$. 
	}
	\label{img:azi_compare}
\end{figure}

We adopted a 2D cylindrical coordinate system, specifically constructing the $ (r, \phi) $ coordinates to configure the grid for the disc. The azimuthal direction $ (\phi) $ was bounded by $ [0, 2\pi] $ to ensure complete coverage of the entire azimuthal domain. For the radial direction $ (r) $, the computational domain is limited from $ [r_{in}, r_{out}] = [0.2, 7.0]\,r_0$. 
To determine the optimal value for $ r_{out} $ and ensure the boundary does not impact results while saving computation time, we explored various outer boundaries ranging from \num{3.1}$r_0$  \citep[][hereafter HA18]{2018MNRASHands} to \num{10}$r_0$. In Fig.~\ref{img:azi_compare}, we compare the azimuthally averaged surface density perturbation, $ \Delta\Sigma_{ave}/\Sigma_{0} $, for two identical models with different $r_{out}$. Furthermore, both models contain almost the same number of cells in the horseshoe region. The behaviour of models remains identical until approximately \num{2.3}$r_0$. However, beyond this point, slight differences emerge due to the proximity of the boundary region to the initial position of the outer planet. To protect against the influence of boundary and damping regions on the torque acting on the outer planet and the overall system evolution, we set $ r_{out} $ to \num{7}$r_0$.

To ensure a proper resolution across the disc, we used a computational grid with evenly spaced azimuthal and logarithmically spaced radial directions. 
We created a computational domain containing $N_{\phi} = 1024$ cells in the azimuthal direction and $N_r = 700$ cells in the radial direction. The resulting cells are approximately square with this grid configuration. We selected the resolution of our simulation to achieve a scale of \num{10} cells per scale height at unit of length, resulting in approximately 4 cells within half of the horseshoe region. According to the study of \citet{2010MNRASPaardekooper}, this number of cells is considered sufficient for resolving the horseshoe region. The chosen resolution constructs a balance between accurately capturing the dynamics of the horseshoe region and computational efficiency.However, we conducted a resolution study in  Sect.~\ref{subsubsec: reso}.

To handle wave reflection in the radial direction near the edges of the domain, we implemented the wave-killing prescription proposed by \citet{2006MNRASDeValBorro}. This technique effectively suppresses wave reflections. The damping zones were placed between [0.2--0.22]$r_0$ and [6.38--7]$r_0$, providing a buffer region where the waves are attenuated without interfering with the main computational domain. However, in the study by HA18, they damped the surface density to zero within wave-killing zones in their \texttt{PLUTO} simulations.

In order to prevent numerical singularities in the evolution of potential gravity near the planets, we chose a smoothing length of $\epsilon = 0.6 H_{p}$, as recommended in the study by \citet{2012A&AMuller}. This smoothing length ensures smooth and stable computations in the vicinity of the planets. To compare, it is noteworthy that HA18 employed a different value, $ \epsilon = 0.4 H_{p} $, in their study.

\begin{figure}[h!]
	\centering
	\includegraphics[width=\columnwidth]{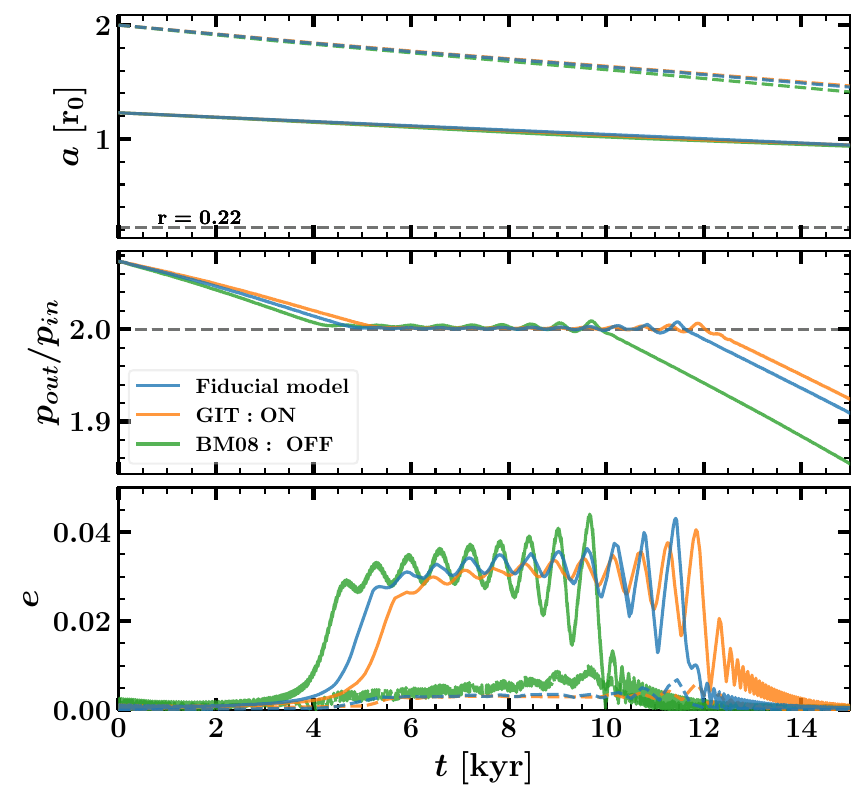}
	\caption{Impact of the GIT and BM08 correction on overstability occurrence. Evolution of the semi-major axis of planets (top), the period ratio of the outer planet to the inner one (middle), and the eccentricity of the inner planet (larger eccentricities) and the outer one(shorter eccentricities; bottom). In the top panel, the solid lines represent the inner planet and the dashed lines the outer planet.
	}
	\label{img:bm08_indir}
\end{figure}

In our models, we considered the indirect-term resulting from the planets' gravity. However, we did not incorporate the gas-indirect term (GIT)\footnote{Explicitly, the parameter \texttt{-DGASINDIRECTTERM} in the \texttt{.opt} file of \texttt{Fargo3d} setup is disabled.} into our analysis. Nevertheless, in Fig.~\ref{img:bm08_indir}, we illustrate the impact of GIT on the occurrence of overstability. Our findings indicate that the GIT does not significantly affect the occurrence of overstability.

In all models, we implemented the torque correction method presented by \citet[][hereafter BM08]{2008Baruteau} to manage the elimination of disc self-gravity. This correction involves subtracting the azimuthally averaged surface density of the disc before calculating the torque exerted on each planet. Additionally, we investigated whether overstability occurs when the BM08 correction is disabled. Figure~\ref{img:bm08_indir} illustrates that overstability indeed occurs in this case. However, we observe that the planets in this setup reach resonance earlier compared to our fiducial model and they spend less time in resonance before breaking it.\\
We summarise the setup in Table~\ref{table: hydro parameters}. 
	\begin{table}[ht] 
		\caption{Parameters of the disc, planets, and grid for all hydrodynamic simulations. The highlighted rows represent the disc parameters of the fiducial model that are changed in the parameter study. 
		}
		\centering 
		\begin{tabular}{l c c} 
			\hline\hline 
			\rule{0pt}{5mm}
			Parameters & Symbol & Values \\ [1ex] 
			\hline
			Surface density slope & $\alpha_{\Sigma}$ & \num{1.0} \\
			Flaring index & $f$ & \num{0.25} \\
			Smoothing length & $\epsilon$ & $0.6 H_{p}$\\
			\hline
		    \rowcolor{lightgray!20} Aspect ratio  &$h_{0}$ & $0.05$ \\ 
		    \rowcolor{lightgray!20} Viscosity parameter & $\alpha_{\nu}$ & $10^{-5}$ \\ 
	    	\rowcolor{lightgray!20} Initial surface density  & $\Sigma_{0}$ & $1.13 \times 10^{-4}\mathrm [\frac{M_{\star}}{r_{0}^{2}}] $ \\ [1ex]
	    	\hline 
  			Inner and outer planets mass & $m_{1}, m_{2}$ & $5 , 10\mathrm [M_{\oplus}]$ \\ 
	  		Semi-major-axis of planets & $ a_{1}, a_{2} $ & $ 1.23, 2.0 \mathrm [r_0]$ \\  [1ex]
  			\hline 
 			Inner and outer boundaries & $r_{in}, r_{out}$ & $ 0.2, 7.0 \mathrm [r_0]$ \\
	  		Spatial resolution & $N_{r}, N_{\phi}$ & $ 700, 1024$\\ [1ex] 
			\hline 
		\end{tabular}
		\label{table: hydro parameters} 
	\end{table}
\section{Results} \label{sec : results}
This section presents the results of our fiducial model, followed by an examination of the impact of viscosity, disc mass, and aspect ratio on the fiducial model. 
We observed (partial)~gaps in some of our models, which had been reported in the recent study by \citet{2021A&AAtaiee}. 
\subsection{Fiducial model} \label{subsec: fiducial}
\begin{figure}[h]
	\centering
	\resizebox{\hsize}{!}{\includegraphics{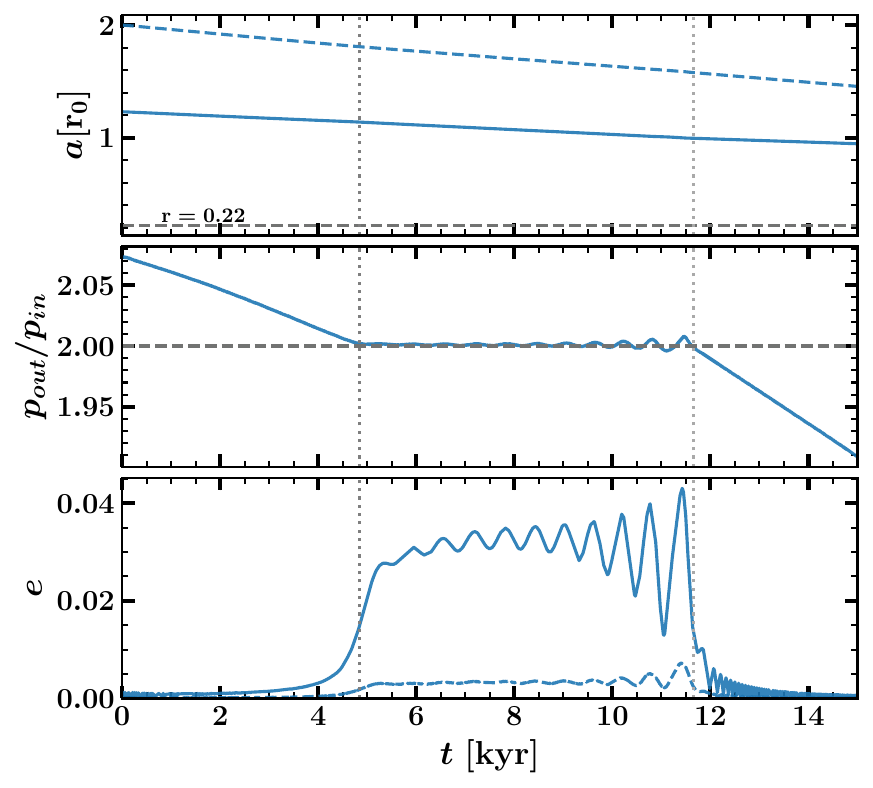}}
	\caption{Time evolution of the planets' orbital properties. From top to bottom: the planets' semi-major axes, the orbital period ratio of the outer planet to the inner one, and the eccentricity of the inner and outer planets. In the top panel, the solid lines represent the inner planet and the dashed lines the outer planet. The two planets are initially located close to a 2:1 resonance. The horizontal dashed line, $r = 0.22$, corresponds to the inner damping radius. The vertical dark and light grey lines represent the time when the planets are in resonance. In our fiducial model, the dark grey line corresponds to $ t = \num{4840} $ years and the light grey line to $ t = \num{11660} $ years.
	}
	\label{img:fiducial model}
\end{figure}

We began by examining the fiducial model, which is characterised by an initial surface density of $ \Sigma_0 = 1.13 \times 10^{-4}~M_{\star}/{r_0}^2 $, a viscous parameter of $\alpha_{\nu} = 10^{-5}$, and an aspect ratio of $h_0 = 0.05$. In this model, two planets are initially placed near 2:1 commensurability, with one planet located at \num{1.23} and the other at \num{2.0}. After about $5000~\mathrm{years}$ of convergent migration, the planets enter into a 2:1 resonance that becomes overstable within approximately $6000~\mathrm{years}$. Then, after this period, the resonance breaks.
Figure~\ref{img:fiducial model} provides an overview of the orbital evolution of the two planets in the fiducial model. In the top panel, we plot the semi-major axes of both planets. The planets migrate inwards, and as the outer planet is more massive, it migrates faster compared to the inner one. In the second panel, we illustrate the period ratio of the outer planet to the inner one. After approximately 5000 years, the planets enter a 2:1 resonance configuration. They remain in this resonant phase for about 6000 years before eventually escaping the resonance. In the last panel, we display the eccentricity evolution of the planets. Initially, the planets have nearly circular orbits, and as they approach the resonant phase, their eccentricities are excited. Then they start to oscillate around an equilibrium value during the resonant period. The amplitude of these oscillations grows over time until the eccentricity is eventually damped back to near zero as they exit the resonance phase. This behaviour of eccentricity is one of the indicators of overstability in the system.

\begin{figure}[h]
	\centering
	\resizebox{\hsize}{!}{\includegraphics{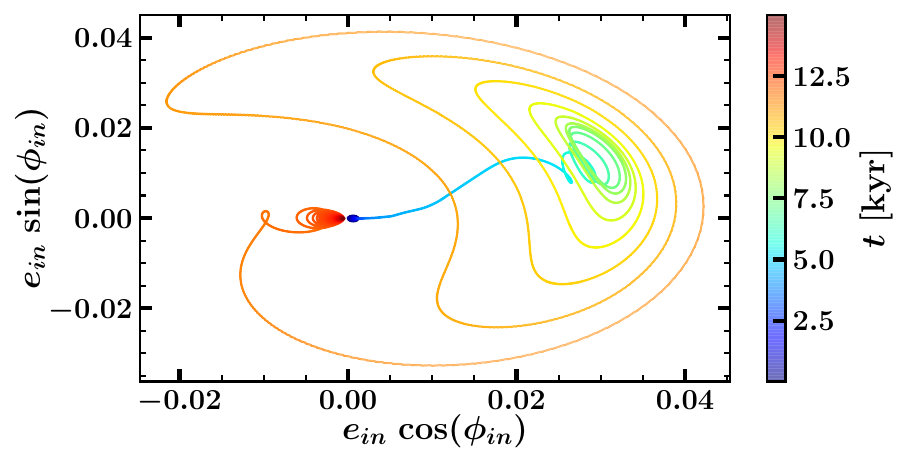}}
	\caption{Phase space of the inner planet for the fiducial model. The colour indicates the time. 
	}
	\label{img:phase_space_fiducial model}
\end{figure}

In Fig.~\ref{img:phase_space_fiducial model} the phase space of the eccentricity and resonance angle of the inner planet is presented. Before the planet's eccentricity is excited, denoted by the blue spot, the eccentricity was initially almost zero, resulting in $ e\sin(\phi) $ and $ e\cos(\phi) $ values of zero. Once the planet enters the resonance phase, both the eccentricity and resonance angle start oscillating around their respective equilibrium values, $ e_{eq} $ and $ \phi_{eq} $. These oscillations are visually represented by the repeating patterns observed in the plot. Over time, the amplitude of these oscillations increases, leading to the size expansion of the patterns in the phase space plot. Upon the planet's exit from the resonance phase, the eccentricity is damped. Simultaneously, the resonance angle shifts between zero and $2\pi$, causing the appearance of circular patterns centred around zero (the reddish part of the curve).
The characteristics of this phase space exhibit similarities to the phase space established by \citet{2014AJGoldreich}, both qualitatively and in terms of the patterns observed. This agreement suggests that the system can be classified as overstable based on the definition provided in Sect.~\ref{subsec:defs}. We leave the investigation of the equilibrium resonance angle to Sect.~\ref{sec: discussion}.

\begin{figure}[h]
	\centering
	\resizebox{\hsize}{!}{\includegraphics{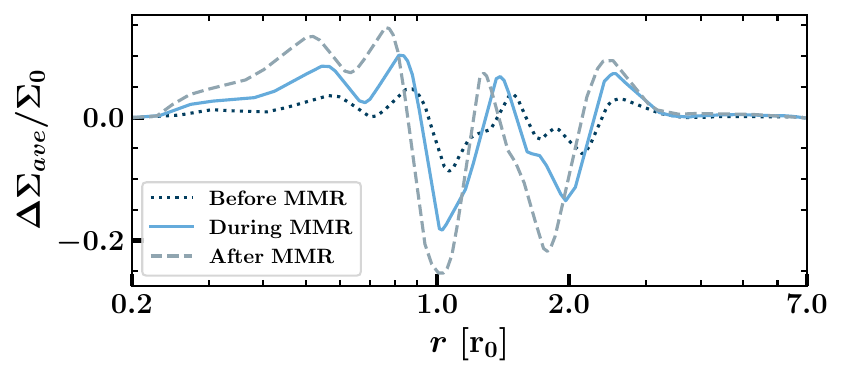}}
	\caption{Azimuthally averaged surface density perturbations before, during, and after resonance capture. The dotted dark grey line shows the average surface density perturbation at t = \num{3000} years, the solid blue at t = \num{7500} years, and the dashed light grey at t = \num{12500} years.}
	\label{img:azi_fiducial}
\end{figure}

In Fig.~\ref{img:azi_fiducial} we illustrate the azimuthally averaged surface density perturbation caused by planets at various time intervals for our fiducial model. 
Both planets carve partial gaps ($\sim 30\%$) in the disc, thereby disrupting its initially smooth structure. This is consistent with findings of \citet{2006Crida} and \citet{2013ApJ.Duffell}, who suggest a higher possibility of gap opening in low-viscosity discs. A detailed analysis and discussion regarding gap opening is presented in Sect.~\ref{sec: discussion}. Additionally, we performed a test for the impact of a disc with opened gaps on overstability occurrence as established in Appendix\ref{app: gap effect}.
\subsection{Parameter study} \label{subsec: parameter}
As part of this study, we investigated the impact of disc parameters, including initial surface density, viscosity, and aspect ratio, on the overstability occurrence of the planetary system. By varying these parameters, we aimed to gain an understanding of how they influence the occurrence and characteristics of the overstability phenomenon in the planetary system.
\subsubsection{The effect of viscosity } \label{subsubsec : viscosity}
We explored the impact of viscosity on the overstability of planetary systems by conducting simulations with various values of the viscosity parameter, $\alpha_{\nu}$, equal to $ [10^{-6}, 10^{-4}, 10^{-3}] $. 

\begin{figure}
	\centering
	\resizebox{\hsize}{!}{\includegraphics{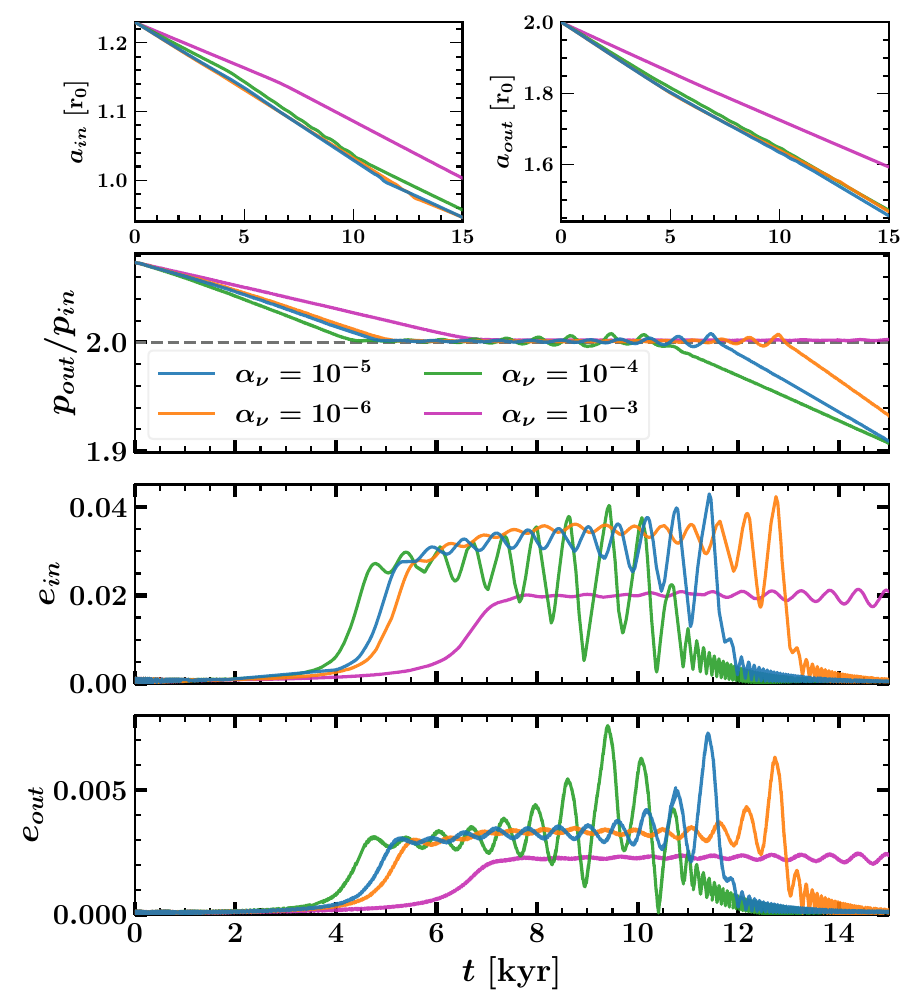}}
	\caption{Influence of viscosity on the evolution of the planetary system for $\alpha_{\nu} = 10^{-6}, 10^{-5}, 10^{-4}$, and  $10^{-3}$, where $10^{-5}$ represents our fiducial model. The figure displays the semi-major axes, the period ratio of planets, and the eccentricity of inner and outer planets from top to bottom. The first row is divided into right and left panels, illustrating the semi-major axes of the inner and outer planets, respectively.
	}
	\label{img:alphaviscosity}
\end{figure}

Figure~\ref{img:alphaviscosity} presents the results of the models with different viscosity. The two panels in the first row illustrate the evolution of the semi-major axis for the inner and outer planets, respectively, from left to right. 
The results indicated that increasing the disc viscosity decreases the migration speed of both planets in the disc. In the lower-viscosity models, the evolution of the semi-major axes of both planets overlaps. 
However, for the model with the highest viscosity in this study $ (\alpha_\nu = 10^{-3}) $, the planets migrated more slowly than the others. This happens because the positive co-rotation torque is larger for this value of viscosity. We discuss this point further in Sect.~\ref{subsubsec: corotation}.

In the second panel of Fig.~\ref{img:alphaviscosity}, which shows the period ratio of the two planets, we observed that in all models, planets entered the 2:1 resonance. In the low-viscosity models, we observed overstable behaviour. However, in the highest-viscosity model in this study ($\alpha_{\nu} = 10^{-3}$), the system remained in resonance. For $\alpha_{\nu} < 10^{-3}$, higher viscosities led to earlier captures into the 2:1 resonance and a shorter duration of the resonance phase. In the $\alpha_{\nu} = 10^{-3} $ model, where planets only captured into resonance without showing signs of overstability, resonance capture occurred very late, and the planets remained in the resonance phase throughout the entire simulation period.

In the last two panels of Fig.~\ref{img:alphaviscosity}, we demonstrate the effect of viscosity on the eccentricity evolution of the planets. When the viscosity parameter, $\alpha_{\nu}$, is less than $10^{-3}$, both planets display typical behaviour of overstable planetary systems. This behaviour is characterised by eccentricity enhancement followed by oscillations around an equilibrium value during the resonance phase. Before the resonance breaking and eccentricity damping, the amplitude of these oscillations grew significantly. The average eccentricity value of the inner planet was approximately ten times higher than that of the outer planet. In the model with $\alpha = 10^{-3}$, the eccentricity remained at an equilibrium value, approximately half of the value observed in other models, but it did not dampen. To investigate this behaviour, we extended the simulation time of the $\alpha = 10^{-3}$ model up to three times the typical simulation duration used for our fiducial model.

\begin{figure}
	\centering
	\resizebox{\hsize}{!}{\includegraphics{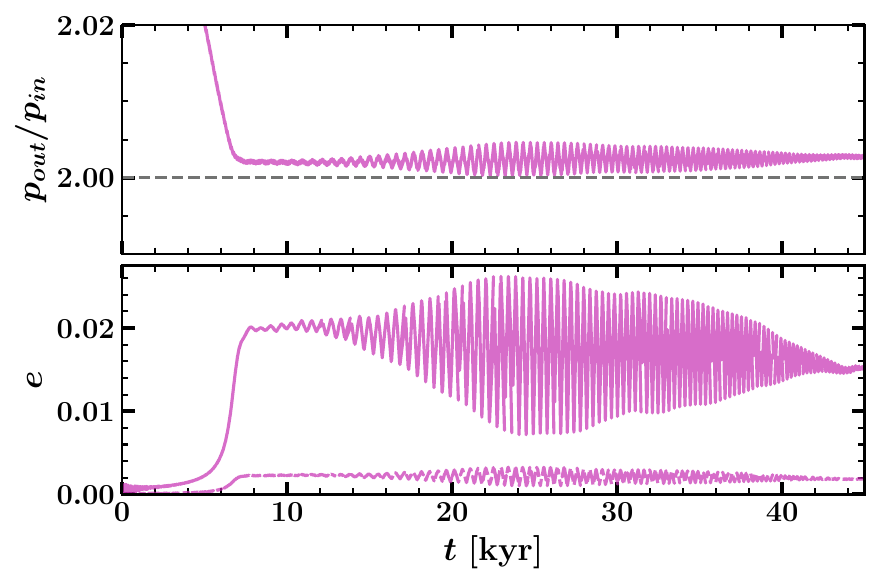}}
	\caption{Period ratio and eccentricity evolution of planets for the model with $\alpha = 10^{-3}$. 
		To track the evolution of the $\alpha = 10^{-3}$ model,  we extended the simulation time up to \num{45000} years. The presence of a limit cycle is evident in the eccentricity evolution, denoted by persisting eccentricity oscillations without damping to zero.
	}
	\label{img:alpha3}
\end{figure}

Figure~\ref{img:alpha3} illustrates the time-extended simulation of the $\alpha_{\nu} = 10^{-3}$ model. The eccentricity exhibited oscillations around an equilibrium value. The amplitude of oscillations initially increased and then decreased. Eventually, the eccentricity stabilised at a constant value without experiencing damping. This phenomenon is known as a limit cycle, where the system's behaviour formed a repetitive pattern over time.

\begin{figure}
	\centering
	\resizebox{\hsize}{!}{\includegraphics{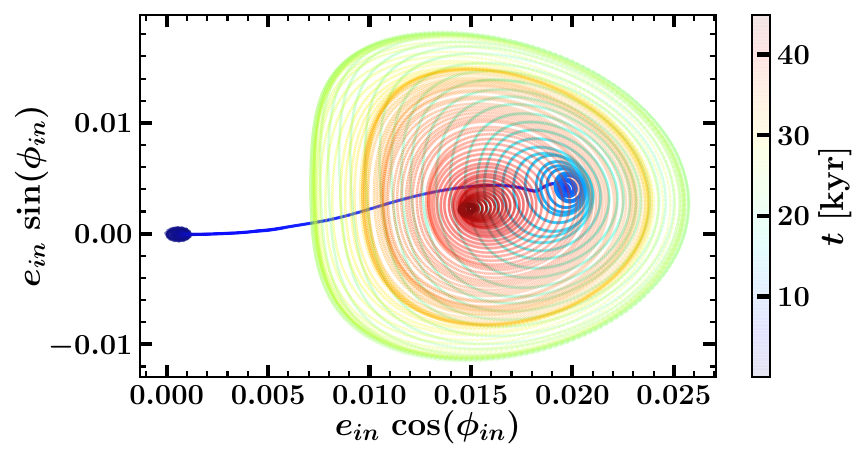}}
	\caption{Phase space of the inner planet for the $\alpha = 10^{-3}$ model. The colour indicates the evolutionary time. The presence of a limit cycle is clearly evident in the phase space, denoted by the repeating oscillations of the inner planet's orbit around a stable equilibrium value. 
	}
	\label{img:Phasespace-alpha3}
\end{figure}

To further confirm the occurrence of a limit cycle in the $\alpha_{\nu} = 10^{-3}$ model, we present its phase space in Fig.~\ref{img:Phasespace-alpha3}. This plot illustrates the excitation of eccentricity from zero, followed by oscillations in both the resonance angle and eccentricity around an equilibrium value. Although the amplitude of these oscillations initially increased and then decreased, the eccentricity and resonance angle consistently remained at the equilibrium value, never damping to zero. The phase space plot clearly confirms the presence of the limit cycle.
\subsubsection{The effect of disc surface density} \label{subsubsec : surface density}
In this section we explore various disc masses, specifically [$\frac{1}{4}$, $\frac{1}{3}$, $\frac{1}{2}$, 2, 3] times $\Sigma_{0}$. We aimed to understand the influence of disc mass variations on the orbital properties of the planetary system.

\begin{figure}
	\centering
	\resizebox{\hsize}{!}{\includegraphics{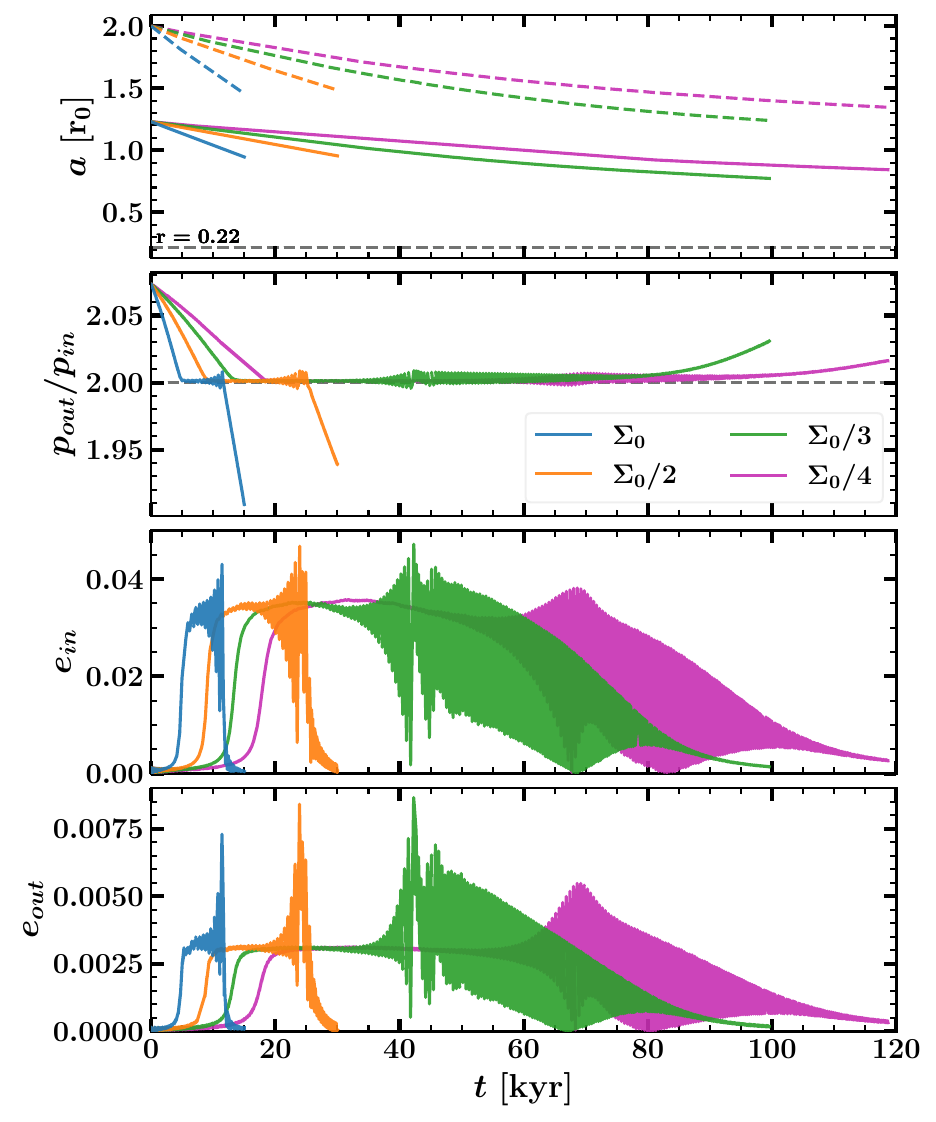}}
	\caption{Impact of a decreasing surface density of the disc on the orbital properties of planets. From top to bottom: Evolution of planets semi-major axes, the period ratio of planets, and the eccentricity of the inner and outer planets. The dashed lines represent the outer planet, while the solid lines represent the inner planet. The dotted line, which corresponds to $r = 0.22$, represents the inner damping radius.
	}
	\label{img:low surface densities}
\end{figure}

In Fig.~\ref{img:low surface densities} we compare the fiducial model with models with initial surface densities of $[\frac{1}{4}, \frac{1}{3}, \frac{1}{2}]\ \rm \Sigma_{0}$.  The first panel, displaying the semi-major axes of the inner and outer planets.
In the second panel, we show the period ratio of the outer planet to the inner one. For $[\frac{1}{2}, \frac{1}{3}, \frac{1}{4}]\ \rm \Sigma_0$ models, the planet enters 2:1 resonance at approximately $t \approx $ [\num{8800}, \num{13200}, \num{17800}] years. As a result, planets embedded in discs with higher surface density are captured into the 2:1 resonance more rapidly. They stay in the resonance phase for a shorter duration, for $[\frac{1}{2}, \frac{1}{3}, \frac{1}{4}]\ \rm \Sigma_0$ models, the residence time in the resonance phase is about  $\Delta t \approx $ [\num{16200}, \num{69200}, \num{81200}] years. 
In the $\frac{1}{2} \Sigma_{0}$ model, the migration was converging similarly to the fiducial model, while for the $[\frac{1}{3}, \frac{1}{4}]\ \rm \Sigma_{0}$ models, the migration diverge after $t \approx $ [\num{82400}, \num{99000}] years. According to \citet{2020kanagawa2}, this divergence can be attributed to the profound gaps created by the planets, estimated to be approximately \num{70} to \num{80} percent of the disc's initial surface density for the outer planet in each model.

In the last two panels, we illustrate the eccentricity evolution of the inner and outer planets. These panels demonstrate the occurrence of overstability in planetary systems with surface density lower than $\Sigma_{0}$. 
As shown in the pioneering studies \citep[e.g.][]{2002-tanaka}, torque is directly proportional to surface density. Torque is also inversely proportional to the migration timescale,  $ \tau_a^{-1} $. Therefore, a reduction in surface density results in a smaller torque and an increased migration timescale.
Our results show that the duration of resonance is longer in lower-surface-density models.

The time at which planets exit overstability differs from the time they exit 2:1 resonance. For $[\frac{1}{3}, \frac{1}{4}] \Sigma_{0}$ models, the duration that planets remain in overstability is about $\Delta t \approx$ [\num{28800}, \num{49200}]. After planets exit overstability, the resonance angle is not restricted to a limited value. It changes between [0, $2\pi$], as illustrated in Figs.~\ref{img:ra0.33} and \ref{img:ra0.25}. 
Thus, the $[\frac{1}{3}, \frac{1}{4}] \Sigma_{0}$ models exhibit overstability until approximately $t \approx$ [\num{42000}, \num{67000}] years. After this period, the resonance angle circulates and no longer oscillates around its equilibrium value. Consequently, the planets are no longer overstable, they simply remain in 2:1 commensurability. After a considerable time, they eventually exit this commensurability.

\begin{figure}
	\centering
	\resizebox{\hsize}{!}{\includegraphics{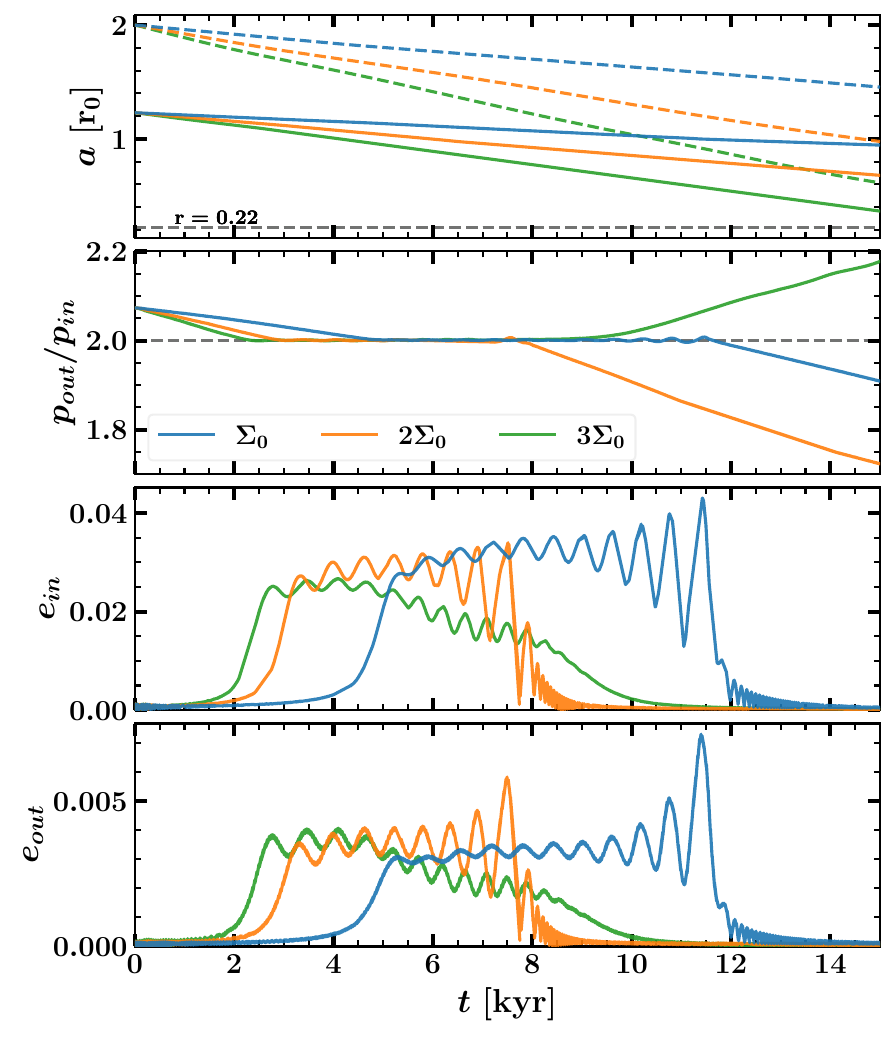}}
	\caption{Impact of an increasing surface density of the disc on the orbital properties of planets. From top to bottom: Evolution of planets semi-major axes, period ratio of planets, and eccentricity of the inner and outer planets. The dashed lines represent the outer planet, while the solid lines represent the inner planet. The dotted line, $r = 0.22$, represents the inner damping radius.
	}
	\label{img:high surface density}
\end{figure}

In Fig.~\ref{img:high surface density}, we present the results of increasing the initial surface density of the disc and its impact on the occurrence of overstability in planetary systems. The first panel illustrates the planets' inward migration, demonstrating the expected linear relationship between surface density and migration rate. As evidenced in the second panel, all models reach the 2:1 resonance. However, contrary to other models, the $3 \Sigma_0$ model diverges upon exiting from resonance. This divergence may be attributed to the wake-planet interaction, which is effective when planets open partial gaps \citep{2013ApJ...778....7B}.

The last two panels depict the eccentricity evolution of the inner and outer planets' orbits. Overstability was evident in the $2 \Sigma_0$ model, whereas in the $3 \Sigma_0$ model, oscillations did not grow around a specific equilibrium value, indicating the absence of overstability. According to the eccentricity evolution in these panels, it is evident that increasing the surface density leads to a decrease in the amplitude of eccentricity oscillations. This indicates that a higher surface density enhances the efficiency of eccentricity damping. In the $3 \Sigma_0$ model, the eccentricity damping efficiency lies within a moderate range. It is not efficient enough to cause a permanent resonance and not low enough to allow the amplitude of oscillations to grow, resulting in overstability. 

As depicted in Figs.~\ref{img:low surface densities} and \ref{img:high surface density}, the variation in the surface density of the protoplanetary disc has a significant impact on the occurrence of overstability, the stability of planetary systems, and their overall structure. In Sect.~\ref{sec: discussion} we explore in detail how variations in surface density influence the architecture of planetary systems.
\subsubsection{The effect of the aspect ratio }\label{subsubsec : aspect ratio}
\begin{figure}
	\centering
	\resizebox{\hsize}{!}{\includegraphics{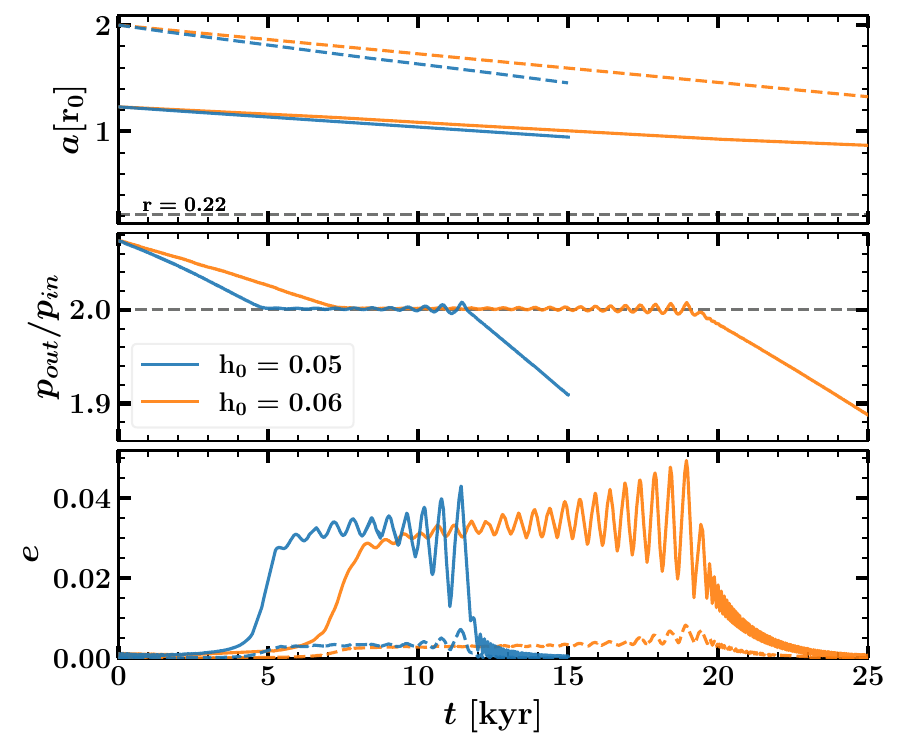}}
	\caption{Impact of increasing the aspect ratio on the evolution of the planetary system. From top to bottom: Evolution of planets semi-major axes, the period ratio of planets, and the eccentricity of the planets. The solid lines represent the inner planet, while the dashed lines represent the outer planet. 
	}
	\label{img:high_h}
\end{figure}

To explore the influence of disc thickness and temperature on the occurrence of overstability, we performed simulations with $h_0 = 0.03, 0.04$, and $0.06$. Figure~\ref{img:high_h} illustrates the impact of increasing the aspect ratio to $h_0 = 0.06$ on the planetary system. In the first panel, we observe that the migration speed slowed with increasing disc thickness, as expected \citep[e.g.][]{2010MNRASPaardekooper}. The second panel shows that the planets reached the 2:1 resonance. However, the capture into resonance occurred later than in the fiducial model, and the planets remained in resonance for a longer period. To fully capture the system's evolution, we extended the simulation runtime. In the last panel, we illustrate the eccentricity evolution of the inner and outer planets. The occurrence of overstability was evident in these plots. Initially, the eccentricity was excited, leading to the growth of oscillations. However, with time, the eccentricity starts to dampen, and eventually, the eccentric orbits turn to circular orbits.

\begin{figure}
	\centering
	\resizebox{\hsize}{!}{\includegraphics{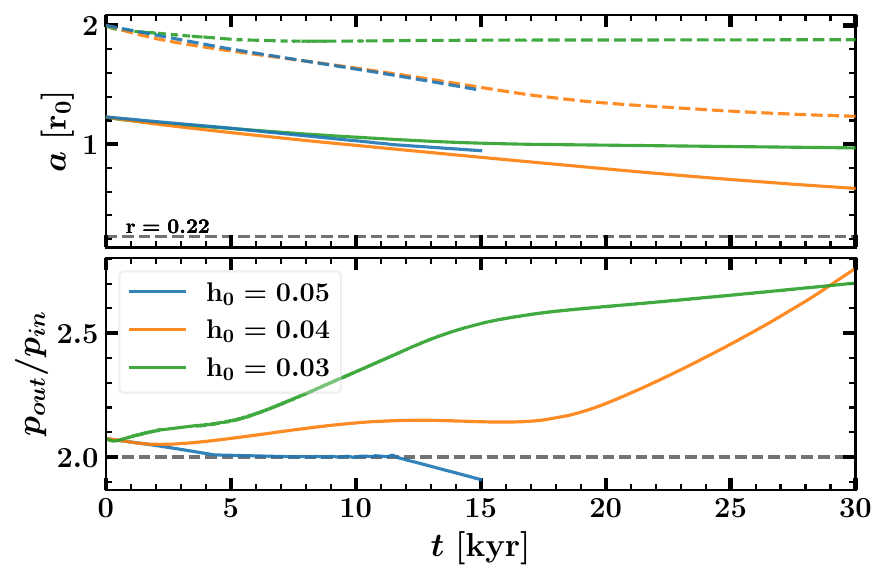}}
	\caption{Influence of a decreasing aspect ratio on the evolution of the planetary system. The first panel shows the evolution of the semi-major axis and the second panel the period ratio of the planets.
	}
	\label{img:low_h}
\end{figure}

In Fig.~\ref{img:low_h} we present the influence of decreasing the aspect ratio of the disc on the evolution of the planetary system.  The first panel displays the inward migration of both planets. Contrary to our expectations from type-I migration, decreasing the aspect ratio did not accelerate migration. Specifically, for the $h_0 = 0.03$ model, the migration speed decreased significantly, while for the $h_0 = 0.04$ model, it remained comparable to our fiducial model. In the second panel, we present the period ratio of the planets. Initially, for the first few years of the simulation, there was an agreement between the period ratio of $h_0 = 0.04$ and the fiducial model. However, after a few years, the migration diverged, and the planets did not even capture into resonance when the aspect ratio was decreased for both models. Instead, they remained in nearly circular orbits throughout the entire simulation without experiencing any overstability. These results indicated that decreasing the aspect ratio of the disc has a significant impact on the migration behaviour and resonance capture of the planets.  Gap openings in these models surely led to these results, as discussed further in Sect.~\ref{sec: discussion}.\\
In Table~\ref{table: parameter study results} we provide a simple overview of the outcomes of our parameter study, indicating whether planet captures into resonance, overstability, or limit cycle occurred.
\begin{table}[ht] 
	\caption{ Concise overview of the outcomes of our parameter study.
	} 
	\centering 
	\begin{tabular}{c c c c} 
		\hline\hline 
		\rule{0pt}{5mm}
		Models & 2:1  & Overstability & Limit \\ [1ex] 
		& Resonance & & Cycle \\ [1ex]
		\hline
		\rule{0pt}{5mm}
		Fiducial Model & \Checkmark & \Checkmark &  \ding{55}\\
		\hline 
		\rule{0pt}{5mm}
		$\alpha_{\nu} = 10^{-6}$ & \Checkmark & \Checkmark & \ding{55} \\
		$\alpha_{\nu} = 10^{-4}$ & \Checkmark & \Checkmark & \ding{55} \\
		$\alpha_{\nu} = 10^{-3}$ & \Checkmark & \ding{55} & \Checkmark \\
		\hline 
		\rule{0pt}{5mm}
		$ \Sigma (r_0) = 1/4\ \Sigma_{0} $ & \Checkmark & \Checkmark & \ding{55} \\
		$ \Sigma (r_0) = 1/3\ \Sigma_{0} $ & \Checkmark & \Checkmark & \ding{55} \\
		$ \Sigma (r_0) = 1/2\ \Sigma_{0} $ & \Checkmark & \Checkmark & \ding{55} \\
		$ \Sigma (r_0) = 2\ \Sigma_{0} $ & \Checkmark & \Checkmark & \ding{55} \\
		$ \Sigma (r_0) = 3\ \Sigma_{0} $ & \Checkmark & \ding{55} & \ding{55} \\ [1ex]
		\hline 
		\rule{0pt}{5mm}
		$ h_{0} = 0.06 $ & \Checkmark & \Checkmark & \ding{55} \\
		$ h_{0} = 0.04 $ & \ding{55} & \ding{55} & \ding{55} \\
		$ h_{0} = 0.03 $ & \ding{55} & \ding{55} & \ding{55} \\ [1ex]
		\hline 
	\end{tabular}
	\label{table: parameter study results} 
\end{table}
\subsection{The effect of numerics}
In this section we explore the impact of numerical parameters on the occurrence of overstability. Initially, we conducted a resolution study to assess how spacial resolution affects the occurrence of overstability. Following this, we compared our results with similar simulations carried out with the \texttt{PLUTO} code.
While recognising the importance of numerical parameters, we observe the occurrence of overstability across the majority of cases.

\subsubsection{Resolution study} \label{subsubsec: reso}
\begin{figure}[h!]
	\centering
	\includegraphics[width=\columnwidth]{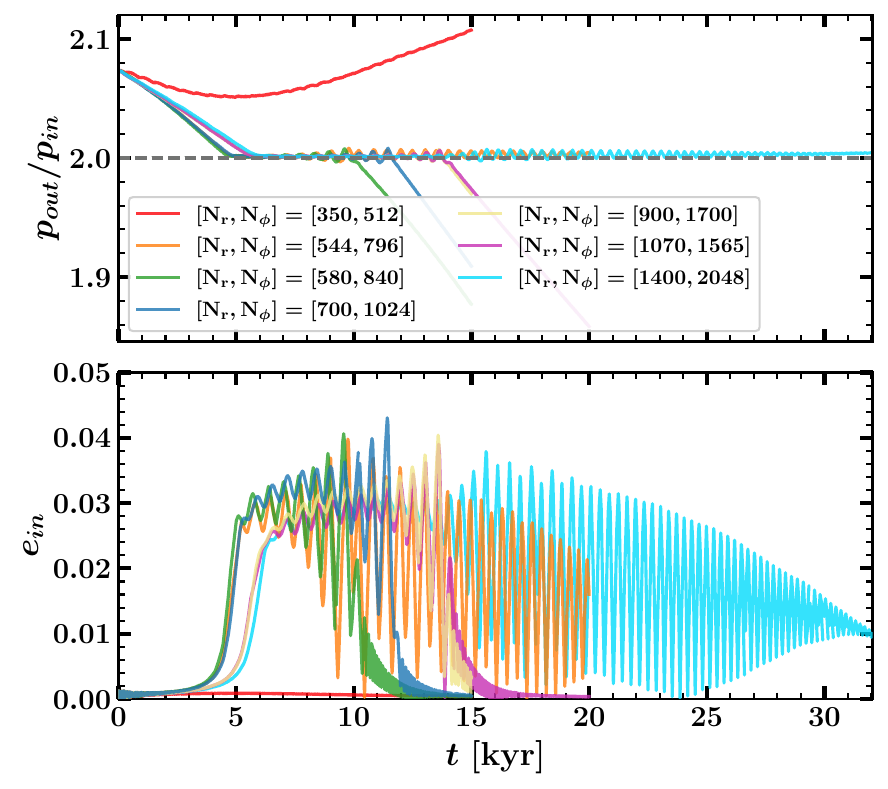}
	\caption{Effect of resolution on overstability occurrence.
	}
	\label{img:reso_compaire}
\end{figure}

To understand how resolution impacts overstability occurrence, we performed simulations at various resolutions while maintaining other parameters consistent with our fiducial model. These resolutions ranged from half to twice that of our fiducial model. The results of these simulations are summarised in Fig.\ref{img:reso_compaire} and Table\ref{table: resolution}.

\begin{table}[ht] 
	\caption{Properties of each resolution and summary of the resolution's impact on overstability occurrence.
	} 
	\centering 
	\begin{tabular}{c c c c c} 
		\hline\hline 
		\rule{0pt}{5mm}
		[$ N_r, N_{\phi} $] & cps  & cp$ x_s $ & 2:1 & Overstability \\ [1ex] 
		& & &  Resonance   & \\ [1ex]
		\hline
		\rule{0pt}{5mm}
		[\num{350}, \num{512}] & \num{5} & \num{2} &  \ding{55} & \ding{55} \\
		
		[\num{544}, \num{796}] & \num{7.6} & \num{3} & \Checkmark & \ding{55} \\
		
		[\num{580}, \num{840}] & \num{8} & \num{3.26} & \Checkmark & \Checkmark \\
		\hline 
		\rule{0pt}{5mm}
		[\num{700}, \num{1024}] & \num{10} & \num{4} & \Checkmark & \Checkmark\\
		\hline 
		\rule{0pt}{5mm}
		[\num{900}, \num{1700}] & \num{12} & \num{5} & \Checkmark & \Checkmark \\
		
		[\num{1070}, \num{1565}]& \num{15} & \num{6} & \Checkmark & \Checkmark \\
		
		[\num{1400}, \num{2048}] & \num{20} & \num{7.8} & \Checkmark & \textbf{?} \\[1ex]
		\hline 
	\end{tabular}
	\label{table: resolution} 
\end{table}

We observe overstability occurrence for $ \rm{cp}x_s = [3.26, 4, 5, 6] $, where $ \rm{cp}x_s$ stands for cells per horseshoe half-width. For $ \rm{cp}x_s =[2, 3] $, the overstability did not occur. However, for $ \rm{cp}x_s =[3]$, the planets reached 2:1 resonance. In $ \rm{cp}x_s = 7.8 $, which is our highest resolution, planets reached the 2:1 resonance, but overstability did not happen during the simulation time, which was about \num{30}kyr. However, from the eccentricity evolution of the planets (and the resonance angle libration, which for the sake of brevity is not shown here), we expect that this model will eventually show overstable behaviour. The difference can be due to how properly the torque on the planet is resolved. It is known that the lowest $ \rm{cp}x_s$ for resolving the horseshoe drag is 4 \citep{2010MNRASPaardekooper}, but we are not aware of how increasing the resolution changes the horseshoe drag on the planet. Further investigation is needed to properly study how the resolution change the torque on the planet that is beyond the scope of this paper.

\subsubsection{Comparison with PLUTO} \label{sec:pluto}
To corroborate our results further, we compared our fiducial model against the numerical hydrodynamics code \texttt{PLUTO} v4.4 \citep{2007Pluto}. By default, \texttt{PLUTO} uses a Godunov-type, finite-volume approach with a second-order space- and time-accurate scheme to advance the hydrodynamics equations. To speed up calculations, and to facilitate a fair comparison to \texttt{FARGO3D}, we also used the FARGO module implemented in \texttt{PLUTO} by \citet{2012Pluto}.

For our fiducial setup, we chose the HLLC Riemann solver \citep{1994Hllc} (a method in computational fluid dynamics for efficiently solving the Riemann problem, incorporating contact waves to capture flow discontinuities and complex phenomena accurately), and second-order reconstruction and time-stepping schemes (\texttt{LINEAR} and \texttt{RK2}, respectively). The planets and star were integrated with an \texttt{RK4} scheme similar to \citet[see Appendix~A therein]{2018Thun}, and the back-reaction of the disc on the star was included. Since we neglected disc self-gravity, we also accounted for the correction to the planetary torques by \citet{2008Baruteau}.

\begin{figure}[h!]
	\centering
	\includegraphics[width=\columnwidth]{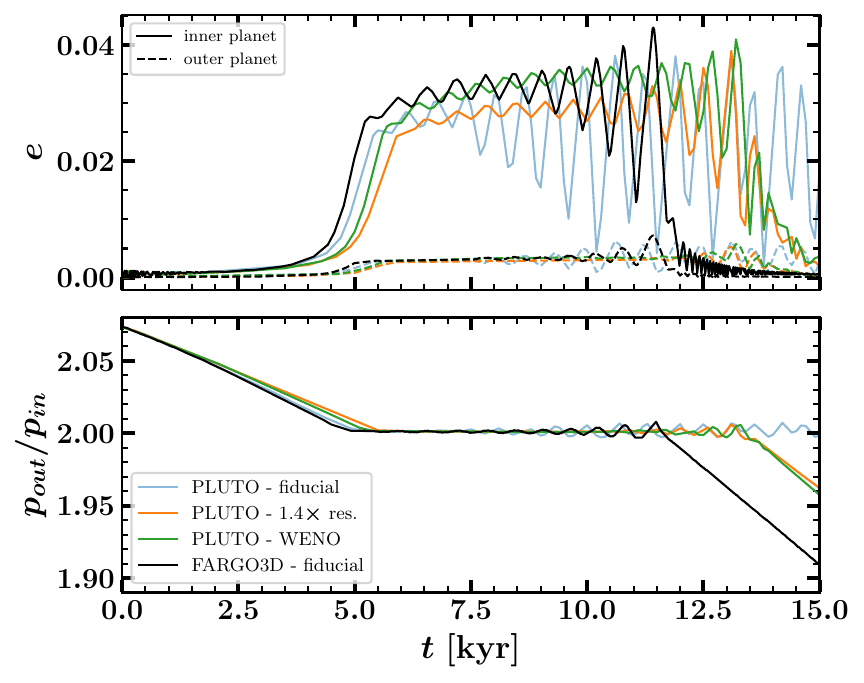}
	\caption{Time evolution of the eccentricity (top) and period ratio (bottom) of the two planets for various configurations of \texttt{PLUTO}, compared to the fiducial \texttt{FARGO3D} setup presented in Table \ref{table: hydro parameters}. The two planets do not break out of resonance in the fiducial \texttt{PLUTO} setup, but a higher resolution or a third-order reconstruction (WENO) both recover the behaviour found by \texttt{FARGO3D}.
	}
	\label{img:pluto}
\end{figure}

Our findings are shown in Fig.~\ref{img:pluto}. While our fiducial model shows an eccentricity growth for both planets similar to \texttt{FARGO3D}, the planets are unable to break the 2:1 resonance. Increasing the resolution by a factor of 1.4 in both directions (for a resolution of $N_r\times N_\phi=990\times1448$ cells), however, allows the planets to break the resonance, indicating that the overstability is operating here. Alternatively, opting for a third-order, weighted essentially non-oscillatory (WENO) reconstruction \citep[][]{2009Weno} at the fiducial resolution also reproduces the overstability, while also matching the amplitude of the eccentricity growth for both planets.

While not shown in the figure, we tested various other combinations of numerical options. We find that, at the fiducial resolution and regardless of numerical setup, the WENO reconstruction is necessary for the planets to break the resonance. At a higher resolution, however, the default numerical setup is sufficient. This suggests that the numerical scheme used by \texttt{PLUTO} is not able to capture the overstability at low resolution, but that the overstability is not a numerical artefact.
This sensitivity to the numerical method between \texttt{PLUTO} and \texttt{FARGO} has also been observed in the context of buoyancy waves \citep{ziampras-etal-2023}, but showed numerical convergence at sufficient resolution.

We note that the self-gravity correction by \citet{2008Baruteau} is essential to trapping the two planets in a 2:1 resonance in the first place.
Without the correction, the planets pass through the 2:1 resonance with only a brief excitation to their eccentricity similar to \citet{2018MNRASHands}, but can still be later captured into 3:2 resonance after around $\sim 8000$~yr.

\section{Discussion} \label{sec: discussion}
\subsection{Equilibrium eccentricity and resonant angle } \label{subsubsec: equilibrium}
\begin{figure}[h!]
	\centering
	\includegraphics[width=\columnwidth]{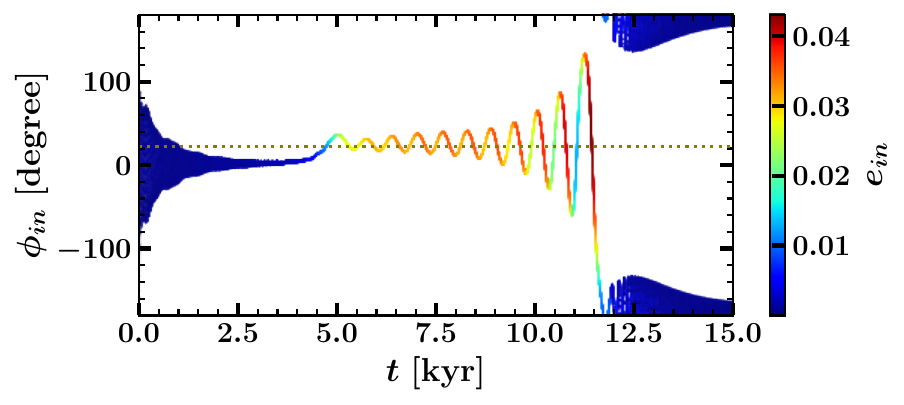}
	\caption{Time evolution of resonant angle for the fiducial model. The colour represents the eccentricity of the inner planet. The horizontal line represents the equilibrium value of the resonant angle during the resonance period of planets, which is $\phi_{eq} = 22.54^{\circ}$.
	}
	\label{img:reso angle}
\end{figure}

The resonant angle and eccentricity oscillate around an equilibrium value, as defined in Sect.~\ref{subsec:defs}. Figure~\ref{img:reso angle} illustrates the evolution of the resonant angle, with the colour scheme indicating changes in the eccentricity of the inner planet. In all of our overstable models, we can estimate the times of entry into and exit from resonance when the eccentricity of the inner planet is approximately $e_{in} = 0.015$. Consequently, by calculating the average of resonant angle in the resonance phase, we determined that $\phi_{eq} = 22.54^{\circ}$ represents the equilibrium value of resonant angle during planetary resonance for our fiducial model.

\begin{figure}[h!]
	\centering
	\includegraphics[width=\columnwidth]{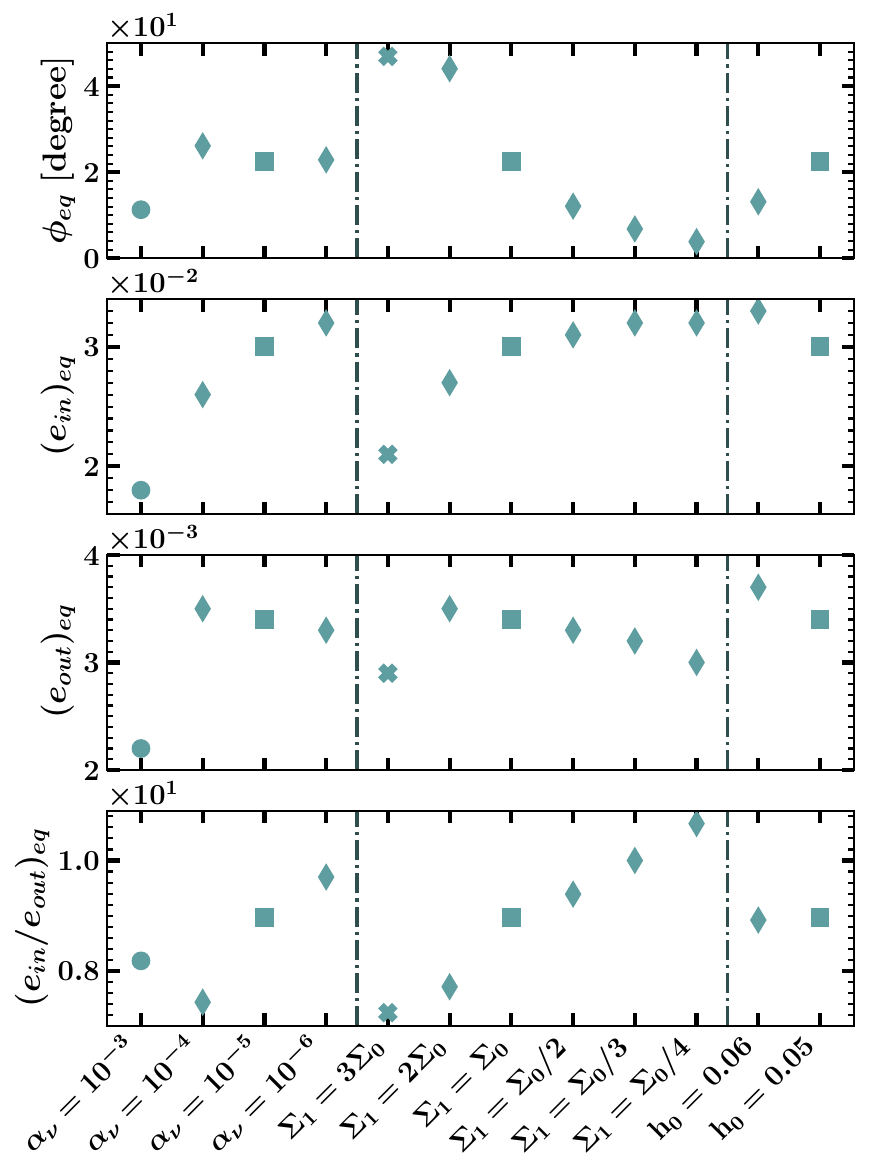}
	\caption{Top to bottom: Equilibrium values for the resonant angle, eccentricities of the inner and outer planets, and the ratio of the planet's eccentricity for the models that captured them in a 2:1 MMR. The viscosity models, initial surface density models, and aspect ratio models are depicted from left to right, separated by vertical lines.  The square, diamond, cross, and circle markers represent the fiducial model, the overstable models, the non-overstable model captured in the 2:1 MMR, and the limit cycle model, respectively. Additionally, $\Sigma_{1} = \Sigma(r_0)$. 
	}
	\label{img:reso all models}
\end{figure}

In Fig.~\ref{img:reso all models} we present the equilibrium values of the resonant angle, the inner and outer planet eccentricities, and the eccentricity ratio for all of our models in 2:1 MMR, regardless of whether they exhibit overstable behaviour. Vertical lines separate different model categories. The left segment comprises viscous models, the middle contains the disc surface density models, and the right segment includes the aspect ratio model.

According to earlier literature \citep[e.g.][]{2002-tanaka}, there exists a simple linear relationship between the eccentricity damping timescale and the migration timescale, $\tau_e = h^2(r) \tau_a$. Additionally, based on the study by \citet{2014AJGoldreich}, the equilibrium resonance angle is given by $\sin\phi_{eq} \propto 1/\tau_e$. Thus, a decrease in $\tau_e$ leads to a reduction in the equilibrium resonance angle.
In viscous models, the equilibrium resonant angle is nearly identical across all three overstable models. As illustrated in Fig.~\ref{img:alphaviscosity}, the overstable viscous models exhibit approximately similar migration rates, resulting in almost identical migration timescales. In contrast, the equilibrium resonant angle value for the limit cycle model is half that of the fiducial model. This discrepancy is reasonable because of the low migration rate associated with the limit cycle model. 
In models investigating surface density, the equilibrium resonant angle decreases as the migration timescale decreases with a reduction in the disc's surface density. According to the equilibrium value of our non-overstable model, we can conclude that models with $\phi_{eq} \approx 50^\circ$ and higher do not experience overstability.
Models examining the aspect ratio reveal a notable decrease in the equilibrium resonant angle and a decreasing migration timescale with an increasing aspect ratio. Therefore, our results align with earlier studies in these cases.

According to the equilibrium eccentricity relation established in \citet{2014AJGoldreich}, $e_{eq} \propto (\tau_e/ \tau_a)^{1/2}$, for our fiducial model, the predicted value of $e_{eq}$ is approximately $0.02$, which deviates from the simulated value of about $0.03$. Similar discrepancies were observed in other models when comparing the $e_{eq}$ values predicted by \citet{2014AJGoldreich} to the simulation results.
These contrasts highlight the need for modifications to the $e_{eq}$ formulation for application in hydrodynamic simulations. Additionally, while eccentricity damping rates and migration rates are given values in N-body simulations and analytical calculations, determining these rates in hydrodynamics involves a more complex process. In hydrodynamic simulations, such as those discussed in this study, eccentricity damping and migration are indeed computed self-consistently through applying the forces from the disc and other planets at every time step.

In \citet{2014AJGoldreich}, a criterion is proposed that if $e_{eq} \geq q_{in}^{1/3}$, the resonance is considered temporary. In our study, where $q_{in}^{1/3} \approx 0.024$, this agrees with our simulations. All our simulations, except the limit cycle and non-overstable models, exhibit equilibrium eccentricity values exceeding the specified threshold. While the limit cycle model seems to be permanently in resonance, the non-overstable model does not exhibit permanent resonance. This suggests that the criterion may be specifically indicative of the presence of overstability in models.

The bottom panel in Fig.~\ref{img:reso all models} illustrates the ratio of the inner planet's eccentricity to the outer one. With the exceptions of the limit cycle model, non-overstable model, and models with $\alpha_{\nu} = 10^{-4}$ and $\Sigma(r_0) = 2\Sigma_{0}$, which closely match the value of $e_{in}/e_{out} \approx 7$ from \citet{2014AJGoldreich}, other models exhibit variations. This suggests that eccentricity and capturing in an exact MMR depend not solely on the planet mass ratio and position but also on disc properties and the interaction between the disc and the planet.

\subsection{The role of co-rotation torque } \label{subsubsec: corotation}

In Fig.~\ref{img:alphaviscosity} we see a considerable trend in the migration rates of models with varying viscosity. We find that the migration rates are nearly identical for models with $\alpha_{\nu}<10^{-3}$. However, for the model with $\alpha_{\nu}=10^{-3}$, the migration rate is remarkably low compared to the others.
To understand the origins of this behaviour, we examined the torque exerted on planets. Our analysis was based on equations found in the studies for Lindblad torques, as well as unsaturated and saturated co-rotation torques \citep[e.g.][]{1980ApJ...241..425G, 2010MNRASPaardekooper,2011paardekooper}.
The scaled Lindblad torque, $\Gamma_{lin}/\Gamma_0 \approx -2.437$, remains unaffected by variations in viscosity, aligning with expectations. In contrast, the co-rotation torque can become saturated with varying viscosity. For $\alpha_{\nu} = [10^{-6},10^{-5},10^{-4},10^{-3}]$, the scaled saturated co-rotation torque is approximately $[0.003, 0.022, 0.141, 0.466]$. Consequently, it becomes crucial to investigate the magnitude of the saturated torque. The value of the unsaturated co-rotation torque remains constant across all viscous models, equal to 2.997. The positive value of the saturated co-rotation torque reduces the absolute value of the total torque to $[2.434, 2.415, 2.293, 1.971]$, resulting in a slower migration of planets, especially in the model with $\alpha = 10^{-3}$.

\subsection{Gap opening } \label{subsec: gap}
In our study we observe that planets create partial gaps around their orbits. In this section we focus on examining these gaps in some of our models, including the fiducial model, the one with $\Sigma (r_0) = \Sigma_{0}/4$, and $h_{0} = 0.03$. 

\begin{figure}[h!]
	\centering
	\includegraphics[width=\columnwidth]{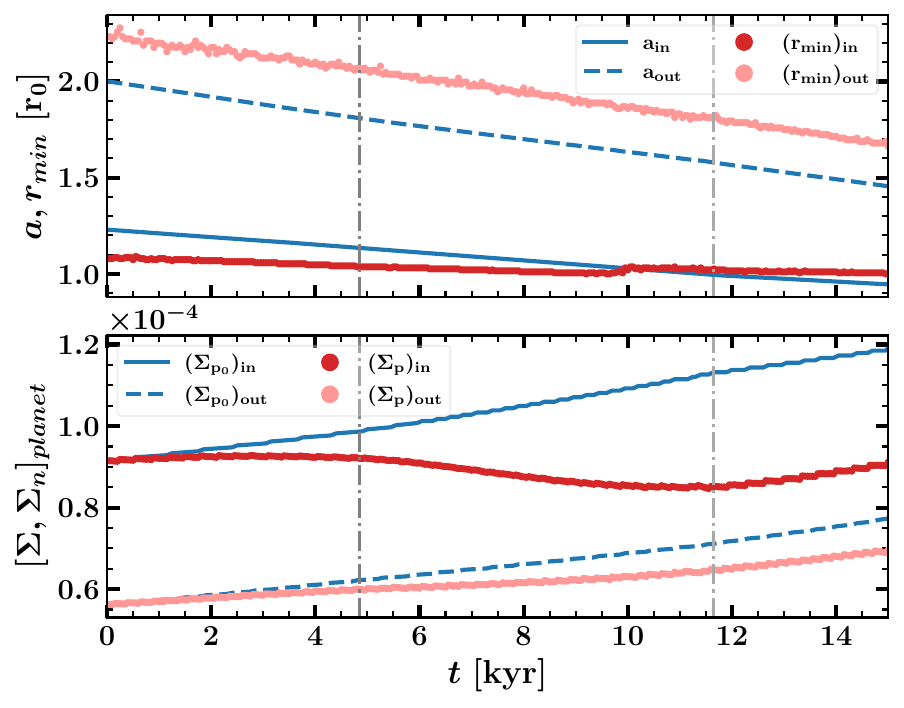}
	\caption{Time evolution of gaps' positions and depths. Top: Time evolution of the semi-major axis and the position at which surface density has its minimum value. Bottom: Time evolution of the surface density in the absence of the planet (disc's power-law surface density) in blue, and the surface density in the position of planets in dark and light red for inner and outer planets.
	}
	\label{img:gap_fiducial}
\end{figure}

Figure~\ref{img:gap_fiducial} depicts the evolution of gaps in our fiducial model. The top panel illustrates the time evolution of the planet's semi-major axes and the position of the bottom of the gaps. We determined the gap's bottom position using an algorithm that identifies a local minimum in the azimuthally averaged 1D profile of surface density around the planet's orbit.
The outer gap, situated beyond the orbit of the outer planet, corresponds to the study by \citet{2020-Kanakawa-ApJ}. Initially, the first gap formed inside the orbit of the inner planet and gradually moved closer to it. However, the gap's position shifted outside the inner planet's orbit around 10,000 years. The bottom panel illustrates the evolution of the disc surface density in the absence (following the power-law relation) and the presence of planets at their positions. The surface density at the initial point, where no gap exists ($\Sigma_n$), increases over time. This increase is attributed to inward migration, leading to movement towards areas of higher surface density. In contrast, in the presence of planets, the surface density at the planet's position ($\Sigma_{pl}$) decreases due to gap formation.

\begin{figure}[h!]
	\centering
	\includegraphics[width=\columnwidth]{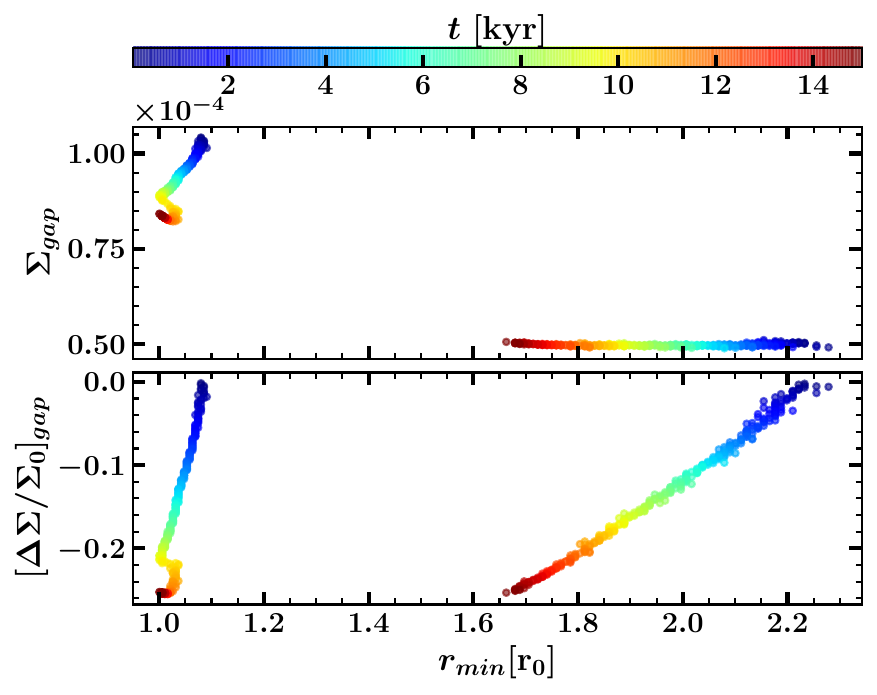}
	\caption{Evolution of the surface density of gaps and its perturbation in terms of positions where surface density has its minimum value. The colour corresponds to time.
	}
	\label{img:gap_scater}
\end{figure}

In Fig.~\ref{img:gap_scater} we display the surface density of gaps and its perturbation, expressed in terms of the position of the bottom of the gap. The colour signifies time. The outer planet's gap is consistently located beyond its orbit. Conversely, the inner planet's gap initially forms within its orbit but gradually moves outwards over time, passing the planet's position.

The inner planet creates a deeper gap compared to the outer one. According to earlier studies \citep[e.g.][]{2006Crida, 2013ApJ.Duffell}, massive planets, low viscosities, and low aspect ratios form deeper gaps. Despite the lower mass of the inner planet, its location in a region with a lower viscosity and a lower aspect ratio leads to the formation of a deeper gap.

\begin{figure}[h!]
	\centering
	\includegraphics[width=\columnwidth]{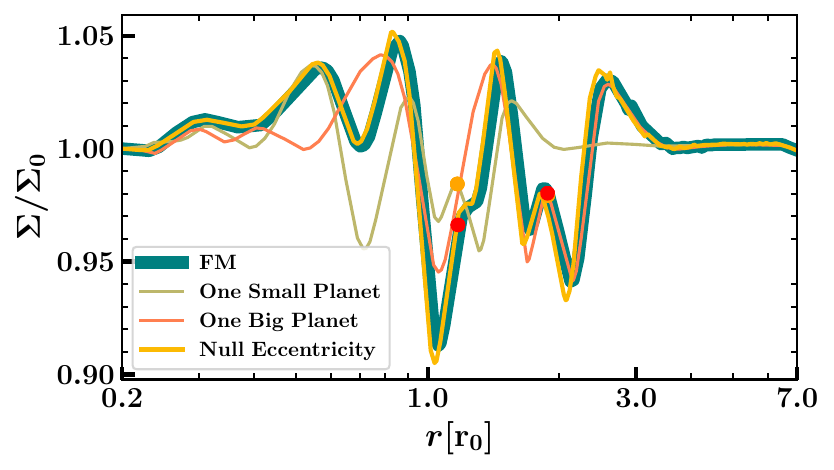}
	\caption{Azimuthal average of disc surface density for multiple models at t = 3000 years. We include our fiducial model (FM), two other models with single planets (one with a low-mass planet and the other with a large-mass planet), and another model identical to the FM but with zero eccentricity. The orange circles represent planets in the single-planet models, while the red circles depict planets in the FM.
	}
	\label{img:azi_fm_one}
\end{figure}

To continue our investigation, we explored the reason behind the placement of the inner planet's gap inside planet's orbit, which differs from the finding by \citet{2020-Kanakawa-ApJ}. Figure~\ref{img:azi_fm_one} illustrates the parameter $ \frac{\Sigma}{\Sigma_0} $, where $\Sigma$ and $\Sigma_0$ denote the values of disc surface density at a specific time point before planetary resonance ($t = 3000$ years) and the initial time ($t = 0$), respectively, for our fiducial model. There are two additional models in this comparative analysis, each with a single planet. One model includes a planet with the mass of the inner planet in our fiducial model, while the other incorporates a planet with the mass of the outer planet in the fiducial model. Both planets in these models are initially positioned in the same locations as those in the fiducial model. In Fig.~\ref{img:azi_fm_one}, it is evident from the single-planet simulations that the gap for both planets lies outside their respective orbits, in agreement with the findings of \citet{2020-Kanakawa-ApJ}. However, in the fiducial model with two planets in resonance, we noticed a distinct behaviour, concerning the gap of the inner planet. This deviation may arise due to the presence of the outer planet. The outer planet, with its notable mass, behaves independently of the inner planet, resulting in its gap opening resembling that of a single planet. On the other hand, the gap opening for the inner planet is different in the models with single and double planets. The gold line shows the gap profile for a test simulation with identical planet locations and evolution time but null eccentricity for the planets. It shows that the eccentricity of the planets are not responsible for the change in the gap profile.

\begin{figure}[h!]
	\centering
	\includegraphics[width=\columnwidth]{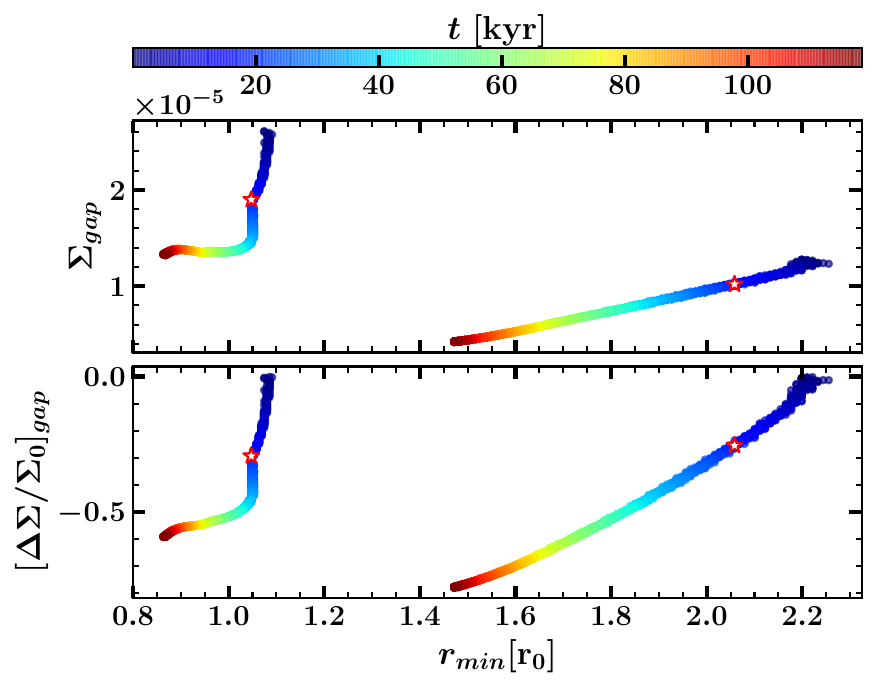}
	\caption{Evolution of gap surface density and its perturbations in terms of the positions of the gap bottoms for the $\Sigma(r_0) = \Sigma_{0}/4$ model. The stars indicate the times when planets enter resonance.
	}
	\label{img:gap_scater_sd0.25}
\end{figure}

After analysing the gap structure of the fiducial model, we shifted our focus to the gap configurations of the $\Sigma(r_0) = \Sigma_{0}/4$ model. Figure~\ref{img:gap_scater_sd0.25} presents the surface density within the gap and its perturbation for the $\Sigma(r_0) = \Sigma_{0}/4$ model. From previous studies \citep[e.g.][]{1980ApJ...241..425G}, we know that decreasing the initial surface density of the disc leads to slow migration.
Figure~\ref{img:low surface densities} highlights that a slower migration speed extends the overstability process. Consequently, planets have more time to deepen the gaps, as seen in Fig.\ref{img:gap_scater_sd0.25}. Contrary to our initial expectation that different surface densities would merely scale the overstability, the gaps deepen due to the longer time of evolution and a low-viscosity parameter, as discussed by \citet{2013ApJ.Duffell}. Similar to the fiducial model, the outer planet's gap is located beyond the planet, whereas the inner gap transitions from inside the orbit to the outside.

\begin{figure*}[h!]
	\centering
	\includegraphics[width=\textwidth]{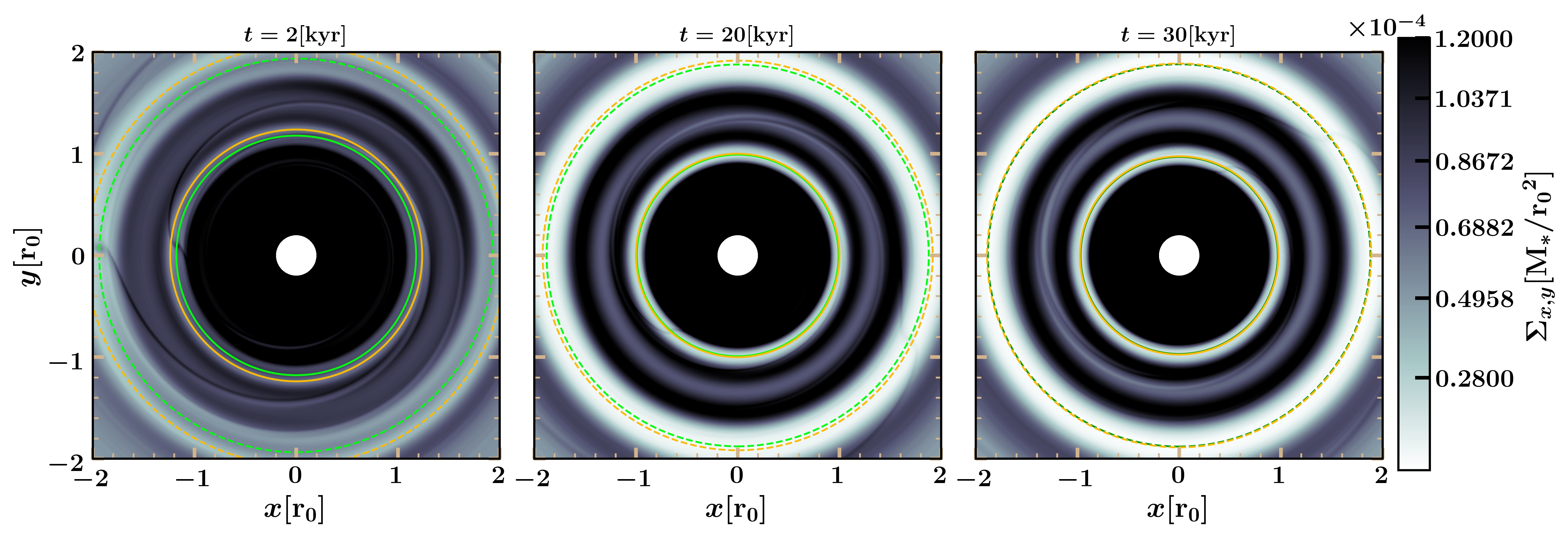}
	\caption{Snapshots of the disc's surface density over time at \num{2000}, \num{20000}, and \num{30000} years for the $h_0 = 0.03$ model. The position of gaps is marked by gold circles, and planets by green circles. 
	}
	\label{img:2d_snapshot_h3}
\end{figure*}

In the following, we investigate the gap structure within the non-resonance $h_0 = 0.03$ model to understand its impact on migration speed. 
According to early studies \citep[e.g.][]{1980ApJ...241..425G}, the scaled torque is proportional to the inverse square of the aspect ratio, $\Gamma_{0} \propto h^{-2}$. Consequently, a planet in a disc with a lower aspect ratio exhibits a higher migration rate. However, our simulations reveal the opposite trend.
In Fig.~\ref{img:2d_snapshot_h3}, we present a 2D snapshot of the disc's surface density at different time points. Even within the initial few thousand years of simulation, planets in this model create notably more profound gaps compared to gaps in the fiducial model, confirming the results from the literature for low aspect ratios \citep[e.g.][]{2006Crida}. These substantial gaps are responsible for the observed slow migration in discs with lower aspect ratios. This is essentially transiting from Type-I to Type-II migration. 

\subsection{Comparison with pure type-I migration} \label{subsec: semi_comparison}
Since the migration torques play a key role in the resonance evolution and occurrence of overstability, we assess in this section the proximity of our planets’ migration rate to pure type-I migration, which is a situation that no gap formed in disc. Given that we have observed gap formation in our planetary systems, we sought to understand how the torque, as calculated for systems with a gap \citep[e.g.][]{2015kanagawa, 2015duffell, 2018Kanagawa}, influences the migration of our planets. For this purpose, we compared our fiducial model's orbital evolution with the study by \citet{2010MNRASPaardekooper}, where no gaps are formed, and with the study by \citet{2018Kanagawa}, accounting for gap formation in the disc \citep[see][Eq. ~(27)]{2018Kanagawa}. It is worth mentioning that we used the same smoothing length as in our hydrodynamic simulations when evaluating Type-I migration. As the co-rotation torque is nearly saturated in this model, the effect of gap opening would be mostly on the Lindblad torque.

\begin{figure}[h!]
	\centering
	\includegraphics[width=\columnwidth]{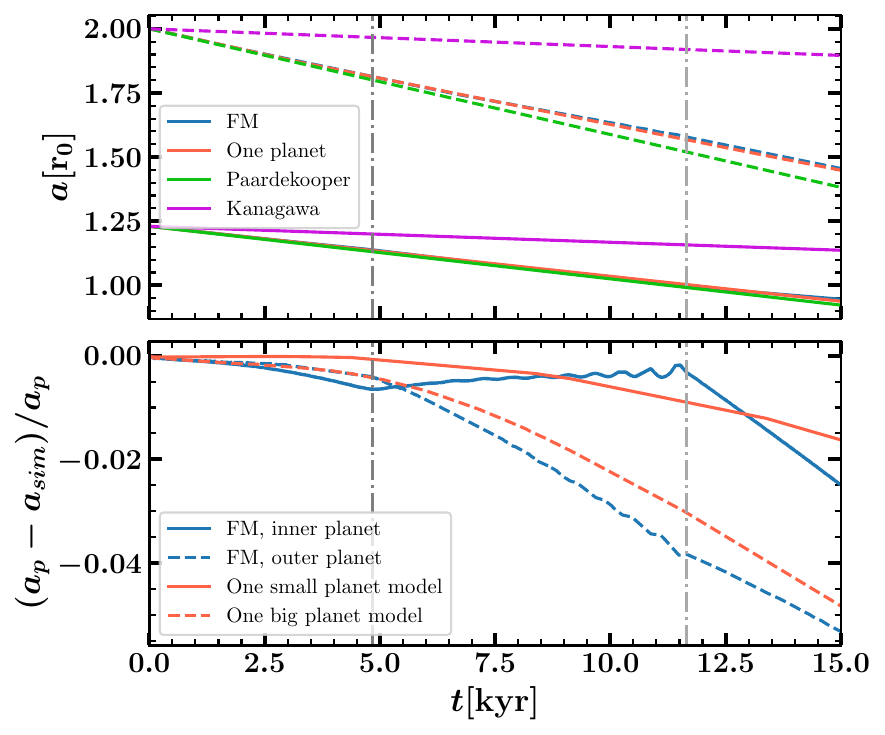}
	\caption{Comparison of semi-major axes obtained through simulation with different theoretical models. Top: Orbital evolution of our FM, along with two semi-theoretical models from \citet[without a gap opening]{2010MNRASPaardekooper} and \citet[with a gap opening]{2018Kanagawa}, alongside two additional simulations using the same disc properties as the FM but focussing on individual planets. Solid lines represent the semi-major axis of the inner planet, while dashed lines represent the outer planet. Bottom: Evolution of the quantity $(a_p - a_{sim})/a_p$, where $a_p$ represents the semi-major axis according to pure type-I migration and $a_{sim}$ denotes the semi-major axis of both the inner and outer planets in our FM, as well as the single small and big planets in our single-planet simulations. The period of resonance is indicated between the dark and light grey time intervals.
	}
	\label{img:semi_compare}
\end{figure}

In top panel of Fig.~\ref{img:semi_compare}, we compare the orbital evolution of both the inner and outer planets obtained from our simulations with the orbital evolution of a planet if the torques on them are calculated according to the prescription in  \citet{2010MNRASPaardekooper} and \citet{2018Kanagawa}. Our simulations agree with the semi-major axes evolution according to pure type-I migration up to a certain point. However, a significant deviation is observed when applying a gap by Kanagawa et al. formula.
This suggests that in torques correction by Kanagawa et al. the depth of the gap for the moderate planetary masses are over-estimated. Concerning the inner planet, its orbital evolution closely follows pure type-I migration until the planet exits resonance. Beyond this point, the differences between the two become apparent. For the outer planet, our simulation diverges from pure type-I migration once the planets reach resonance.

To investigate the difference between our fiducial model and pure type-I migration, we performed two additional simulations with only a single planet in each run, with one simulation for the inner low-mass planet (solid line) and one for the outer high-mass planet (dashed line), in red in top panel of Fig.~\ref{img:semi_compare}. Comparing these single-planet simulations shows a closer agreement with our fiducial model, suggesting that gap formation has minor effect on the migration rate of the planet.

To further assess this difference, we analysed the evolution of the quantity $(a_p - a_{sim}) / a_p$, where $a_{sim}$ and $a_p$ represent the semi-major axis of the planet in our simulations and pure type-I migration. The bottom panel of Fig.~\ref{img:semi_compare}, indicates that the discrepancy between our fiducial model and pure type-I migration is mainly attributed to resonance and gravitational interactions to a minimal extent.

\begin{figure}[h!]
	\centering
	\includegraphics[width=\columnwidth]{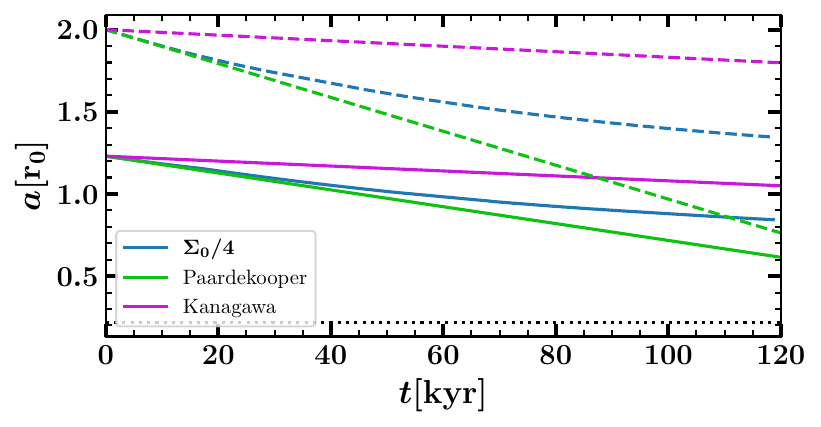}
	\caption{Same as the top panel of Fig.~\ref{img:semi_compare} but for the $\Sigma (r_0) = \Sigma_{0}/4$ model.
	}
	\label{img:semi0.25}
\end{figure}

In the following analysis, we aim to assess whether the model with $\Sigma(r_0) = \frac{\Sigma_0}{4}$, characterised by the formation of a deep gap as depicted in Fig.\ref{img:gap_scater_sd0.25}, agrees with the migration rate predicted by Kanagawa et al., for discs exhibiting significant gaps.
Figure~\ref{img:semi0.25} presents a comparison between the semi-major axis of this model, pure type-I migration, and migration with gap opening. A substantial deviation between the simulation and the Kanagawa et al. theory is observed, particularly in the early stages. This suggests that the long-term evolution of the system led to the formation of these pronounced gaps \citep[e.g.][]{2013ApJ.Duffell}, and the deviation from pure type-I migration in later times indicates that type-I migration in this model is no longer pure.

\begin{figure}[h!]
	\centering
	\includegraphics[width=\columnwidth]{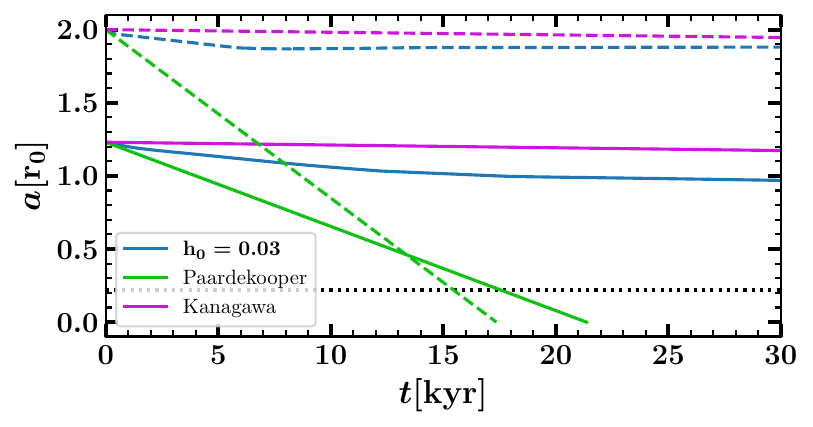}
	\caption{Same as the top panel of Fig.~\ref{img:semi_compare} but for the $h_{0} = 0.03$ model.
	}
	\label{img:semih3}
\end{figure}

In the subsequent analysis, we compare the simulation results of the $h = 0.03$ model with semi-analytical predictions. As illustrated in Fig.~\ref{img:semih3}, a remarkable agreement is observed between the semi-major axis of the $h = 0.03$ model and the prediction presented by Kanagawa et al., for discs featuring significant gaps. This close correspondence strongly supports the presence of a deep gap in the structure of this disc. Despite the planets having a moderate mass, the migration behaviour closely resembles type-II migration.
The figure indicates that the semi-major axis, that predicted by Kanagawa et al., exhibits almost no migration. This is attributed to the Kanagawa et al. expectation that $\frac{\Sigma_{\text{min}}}{\Sigma_0}$ equals \num{0.03} for $h_0 = 0.03$, signifying a gap with a depth of  \num{97} percent. In such a deep gap, the planets essentially orbit around a fixed semi-major axis.
     
\section{Conclusion} \label{sec: conclusion}
We have investigated the overstability in a 2:1 resonant planetary system, considering two planets with moderate masses of \num{5} and \num{10}~$ M_{\oplus} $, using hydrodynamic simulations.
Through an organised parameter study, we found that disc properties, especially mass and viscosity, play a key role in the occurrence of overstability. Low-viscosity discs exhibited overstability, while a high viscosity led to limit cycle phenomena. Variations in the disc mass had a profound impact, with low-mass discs consistently displaying overstability. In these models, planets have enough time to open deeper gaps compared to the higher-surface-density models. For lower-aspect-ratio discs (e.g. $h = 0.03$), the planets formed very deep gaps and underwent type-II migration.
Our study takes a step closer to reality and shows that overstability is not merely a theoretical construct, especially in low-mass and low-viscosity discs: it can aid in forming the packed resonant system. We would like to emphasise that our results are particularly relevant for planets undergoing convergent migration and crossing the 2:1 MMR at approximately one astronomical unit separations from the star.\\
In a future study, we aim to calculate migration and eccentricity timescales in our hydrodynamic simulations and compare the results with theoretical calculations of previous studies carried out using our hydrodynamical simulations.\\
This study pushes us closer to understanding the complex interaction between planetary migration and disc properties. Overstability, previously restricted to theoretical domains, is now on the threshold of empirical confirmation.
	
\begin{acknowledgements}
		This work was supported by Ferdowsi University of Mashhad under Grant No. 3/58699 (1401/11/30).
		The authors acknowledge support by the High Performance and Cloud Computing Group at the Zentrum für Datenverarbeitung of the University of Tübingen, the state of Baden-Württemberg through bwHPC and the German Research Foundation (DFG) through grant no INST 37/935-1 FUGG. 
		The authors thank Cornelis Dullemond and Gabriele Pichierri for their helpful discussions. In heartfelt remembrance, ZA expresses her deep gratitude to Willy Kley, whose unwavering guidance, support, and encouragement have deeply influenced her life journey. 
		We would like to thank the referee O.~Chrenko for his careful comments which helps to improve the manuscript.
		
\end{acknowledgements}

\bibliography{overstable}
\bibliographystyle{aa}


\clearpage
\appendix
\begin{appendix}
	
		\section{Impact of a disc with opened gaps on overstability occurrence} \label{app: gap effect}
\begin{figure}[h!]
			\centering
			\includegraphics[width=\columnwidth]{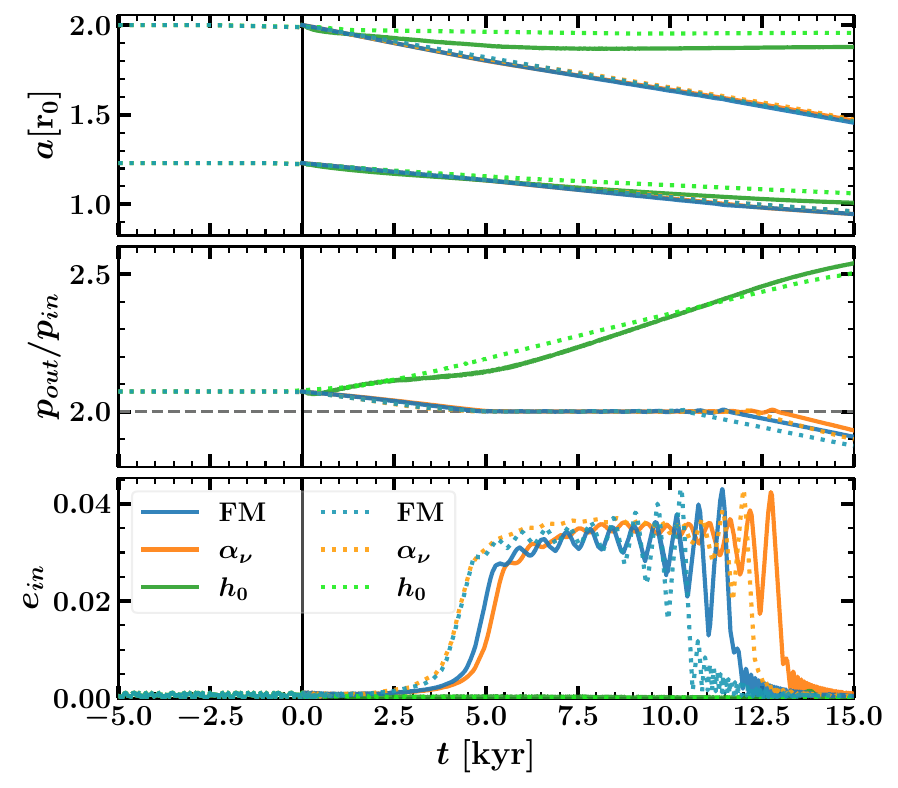}
			\caption{Impact of the disc containing opened gaps on the occurrence of overstability. The solid lines show the models that the evolution of the system is together with gap formation. The dotted lines illustrate the impact of the disc with opened gaps on the system.
			}
			\label{img: gap effect}
		\end{figure}

To examine the influence of opened gaps on overstability occurrence, we conducted three simulations using different models, including our fiducial model, the $\alpha_{\nu} = 10^{-6}$ model, and the $h_0 = 0.03$ model. In these simulations, we initially allowed the system to evolve for 5000 years without migration. Subsequently, after the planets opened gaps, we enabled migration for the rest of the simulation time. The results are depicted in Fig.~\ref{img: gap effect}.
It is apparent that for models experiencing overstability simultaneously with gap formation, the overstability reoccurred, albeit with slight variations in resonance braking time. Additionally, in the model where overstability did not occur together with gap opening, we did not observe overstability again. 

		\section{Resonance angle of low-surface-density discs} \label{app: A}
		\begin{figure}[h!]
			\centering
			\includegraphics[width=\columnwidth]{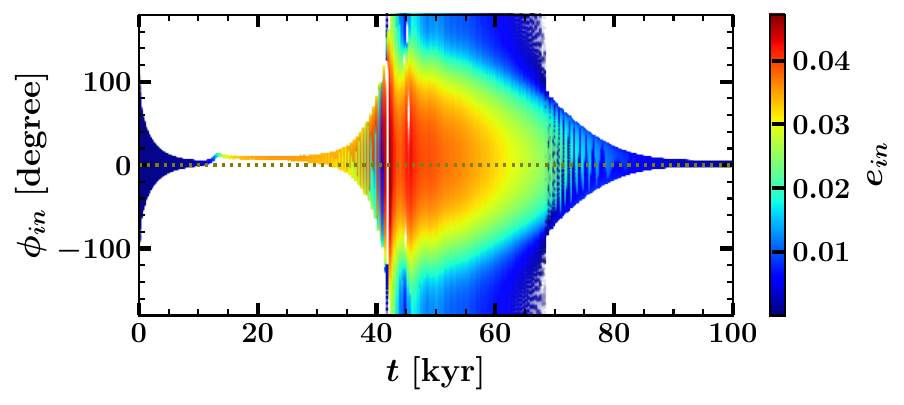}
			\caption{ Resonance angle evolution for the $ \Sigma({r_0}) = \frac{1}{3} \Sigma_{0}$ model. The colour indicates the eccentricity of the inner planet.
			}
			\label{img:ra0.33}
		\end{figure}
	
		\begin{figure}[h!]
			\centering
			\includegraphics[width=\columnwidth]{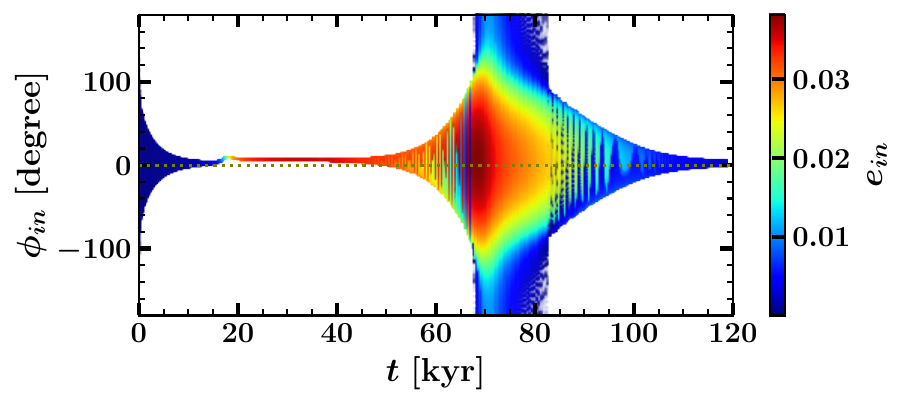}
			\caption{Resonance angle evolution for the $ \Sigma({r_0}) = \frac{1}{4} \Sigma_{0}$ model. The colour indicates the eccentricity of the inner planet.
			}
			\label{img:ra0.25}
		\end{figure}
	
\end{appendix}

\end{document}